\documentclass[twocolumn]{aastex62}

\usepackage{color}
\usepackage{natbib}
\usepackage{graphicx}
\usepackage{hyperref}
\usepackage{cleveref}

\accepted{September 10, 2020}
\submitjournal{ApJ}

\shorttitle{Auroral Emission from L Dwarfs}
\shortauthors{Richey-Yowell et al. (2020)}

\begin{document}

\title{On the Correlation between L Dwarf Optical and Infrared Variability and Radio Aurorae}

\author{Tyler Richey-Yowell}
\affil{School of Earth and Space Exploration, Arizona State University, Tempe, AZ 85281, USA}
\email{tricheyy@asu.edu}

\author{Melodie M. Kao}
\affil{School of Earth and Space Exploration, Arizona State University, Tempe, AZ 85281, USA}
\affil{NASA Hubble Postdoctoral Fellow}

\author{J. Sebastian Pineda}
\affil{University of Colorado Boulder, Laboratory for Atmospheric and Space Physics, 3665 Discovery Drive, Boulder CO, 80303, USA}

\author{Evgenya L. Shkolnik}
\affil{School of Earth and Space Exploration, Arizona State University, Tempe, AZ 85281, USA}

\author{Gregg Hallinan}
\affil{California Institute of Technology, Department of Astronomy, 1200 E. California Avenue, Pasadena CA, 91125, USA}

\begin{abstract}

Photometric variability attributed to cloud phenomena is common in L/T transition brown dwarfs. Recent studies show that such variability may also trace aurorae, suggesting that localized magnetic heating may contribute to observed brown dwarf photometric variability. We assess this potential correlation with a survey of 17 photometrically variable brown dwarfs using the Karl G. Jansky Very Large Array (VLA) at 4 -- 8 GHz. We detect quiescent and highly circularly polarized flaring emission from one source, 2MASS J17502484-0016151, which we attribute to auroral electron cyclotron maser emission. The detected auroral emission extends throughout the frequency band at $\sim$5 -- 25$\sigma$, and we do not detect evidence of a cutoff.  Our detection confirms that 2MASS J17502484-0016151 hosts a magnetic field strength of $\geq$2.9 kG, similar to those of other radio-bright ultracool dwarfs. We show that H$\alpha$ emission continues to be an accurate tracer of auroral activity in brown dwarfs. Supplementing our study with data from the literature, we calculate the occurrence rates of quiescent emission in L dwarfs with low- and high-amplitude variability and conclude that high amplitude O/IR variability does not trace radio magnetic activity in L dwarfs. 

\end{abstract}


\section{Introduction}\label{sec:intro}

Even before the first confirmed discovery of a brown dwarf by \citet{Nakajima1995}, theoretical models of brown dwarfs have long been concerned with the interpretation of clouds in their atmospheres \citep[e.g.][]{Lunine1989}. Prior to the development of real cloud treatments, cloudless models were used to trace the brown dwarf spectral sequence. While some studies argue that cloudless models are still applicable \citep[e.g.][]{Tremblin2015, Tremblin2016}, many others have argued that clouds are ubiquitous within brown dwarf atmospheres and play a key role in our understanding of the evolution of brown dwarfs as they cool throughout their lifetimes. For instance, the transition between L and T spectral types occurs when iron, silicates, and metal oxide compounds condense and begin raining out of the atmosphere \citep{allard2001, tsuji2002}. The remaining cloud coverage is expected to be patchy, which may be the primary source of photometric variability in the optical and infrared \citep[e.g.][]{Apai2013, Radigan2013}. Numerous ground- and space-based studies demonstrate that most ($>$50\%) of brown dwarfs exhibit optical and infrared (O/IR) variability \citep[e.g.][]{Radigan2014, Heinze2015, Metchev2015}.  Such variability can be periodic or irregular \citep{bailer-jones2001, Koen2005, Metchev2015}. Because the atmospheres of brown dwarfs are expected to be neutral from their cool ($<$2000 K) temperatures, \citet{mohanty2002} and \citet{Radigan2013} proposed silicate clouds as the source of the observed variability. 

In the last decade, the discovery that brown dwarfs emit aurorae underscores the possibility that localized magnetic heating due to the energy deposition from the auroral currents may also play a role in brown dwarf variability. \citet{hallinan2007} confirmed that brown dwarf radio flares, first detected on LP 944-20 by \citet{berger2001}, are emitted via the electron cyclotron maser instability (ECMI).  ECMI is also the source of Jupiter's radio aurorae \citep{zarka1992}, and  \citet{hallinan2015} argued that a single magnetospheric current could cause the simultaneous periodic optical and radio variability observed from the brown dwarf LSR J1835+3259.  Soon thereafter, \citet{Kao2016} demonstrated that tracers of Jovian aurorae such as H$\alpha$ emission \citep[e.g.][]{Clary1975} and infrared photometric variability \citep[e.g.][]{Caldwell1980} also appear to be correlated with brown dwarf radio aurorae, further evidence that brown dwarf radio emission was auroral in nature.  Finally, \citet{pineda2017} showed that brown dwarf radio and H$\alpha$ luminosities are correlated.  This suggested that, despite the lack of global coronal heating indicated by the sharp drop-off in X-ray luminosities for brown dwarfs \citep{williams2014}, radio and H$\alpha$ emission from brown dwarfs trace the same current systems.

So far though, models interpreting observed photometric variability in L/T transition brown dwarfs do not take into account the role of localized magnetic heating from auroral currents, as this mechanism and other inhomogeneous surface features cannot be reproduced with current 1-D cloud models \citep{biller2017}. However, spectral models of T-dwarf atmospheres show that localized atmospheric heating can result in excess flux at at 1 -- 10 $\mu$m \citep{morley2014}. Similarly, \citet{robinson2014} show that periodic heating perturbations may produce flux variations on the order of 1 -- 3\% on timescales of both hours and days, including temporal phase shifts of the maximum flux observed at different wavelengths. Thus, thermal influences may additionally contribute to the photometric variability seen on brown dwarfs. Energy deposition from the auroral currents impacting the atmosphere may be one such source of thermal influence.

\citet{hallinan2015} and later \citet{Kao2016} suggested that the inferred non-thermal electron beams traversing the magnetospheres of these brown dwarfs implied by auroral detections could cause spot-heating at the base of these electron beams in the upper atmospheres of brown dwarfs. These types of interactions are readily seen in Jupiter's \citep[e.g.][]{Drossart1989}, Saturn's \citep[e.g.][]{Geballe1993}, and Uranus' aurorae \citep[e.g.][]{Trafton1993}. In the case of Jupiter, both models and observations have demonstrated that magnetosphere-ionosphere coupling drive the thermal profile of the atmosphere. Using a fully 3-D Jupiter Thermospheric General Circulation Model, \citet{Bougher2005} showed that both moderate auroral particle and Joule heating are necessary to recreate observed temperatures over a range of latitudes above the homopause. Recent observations in the infrared by \citet{sinclair2019} have demonstrated that the brightness temperatures of Jupiter's poles increase by several Kelvin ($\sim$25\%) in a matter of days with in an increase in auroral power, in this case due to the solar wind. These observations suggest that the auroral heating on Jupiter may occur even as deep as the upper stratosphere (10 -- 1 $\mu$bar). 

Targeted searches have shown that there may indeed be a connection between these auroral features and the photometric variability. \citet{Harding2013} observed six objects for which auroral radio emission was detected and found that five displayed infrared variability associated with the radio-measured rotation period of the brown dwarf, with the sixth showing a marginal detection. \citet{hallinan2015} demonstrated that a single auroral feature can explain optical photometric variability at different bands that is both in and out of phase for the radio aurora emitting M9.5 dwarf LSR J1835+3259. Similarly, \citet{Kao2016} observed six additional late L and T dwarfs known to exhibit H$\alpha$ emission and/or O/IR variables and found five of six to be auroral radio sources, demonstrating that there may also be a connection from radio emission to H$\alpha$ emission and/or O/IR variability. 

Further characterizing the possible overlap between observational markers of magnetism and clouds on brown dwarfs is imperative for accurately modeling brown dwarf cloud characteristics. We present a search for radio emission indicative of magnetism in a radio survey of 17 brown dwarfs using the the Karl G. Jansky Very Large Array (VLA) from 4 -- 8 GHz (\S\ref{sec:observations}). Our targets are late L dwarfs in the transition region from L to T spectral types, where cooling temperatures cause clouds to precipitate out and result in patchy cloud coverage with strong O/IR variability amplitudes (0.5 - 26\%; \S\ref{sec:targets}). Our target sample allows us to statistically constrain the presence of localized atmospheric heating to observed photometric variability attributed to cloud phenomena (\S\ref{sec:results}, \S\ref{sec:discussion}).

\begin{deluxetable*}{l c c c c c c c c c c}[t]
\centering
\tiny
\tabletypesize{\tiny}
\tablecaption{\normalsize{Target information for the 17 L dwarfs in our sample. Blanks indicate no measurement.}\label{tab:target_info}} 

\tablehead{
\colhead{Object Name} \vspace{-0.3cm} 	&	 \colhead{Abbrev.} 	&	 \colhead{SpT} 	&	\colhead{Ref.}	&	 \colhead{Distance} 			&	\colhead{Ref.}	&	 \colhead{$\mu_\alpha$ cos $\delta$} 			&	 \colhead{$\mu_\delta$}			&	\colhead{Ref.} & \colhead{$\text{log}(L_{\text{H}\alpha} / L_{\text{bol}})$} &	\colhead{Ref.}	\\
\colhead{} 	&	 \colhead{Name} 	&	 \colhead{} 	&		&	 \colhead{[pc]} 			&		&	 \colhead{[mas yr$^{-1}$]} 			&	 \colhead{[mas yr$^{-1}$]}			&	& &	}			
\startdata											
2MASSI J0030300-145033	&	2M0030-14	&	L6.5	&	1	&	26.72	$\pm$	3.21	&	10	&	245	$\pm$	4	&	-28	$\pm$	2	&	13	&	$<$	−5.04	&	18	\\
2MASSI J0103320+193536	&	2M0103+19	&	L6	&	2	&	21.32	$\pm$	3.46	&	10	&	305	$\pm$	17	&	35	$\pm$	14	&	14	&	$<$	−5.96	&	18	\\
2MASS J01075242+0041563	&	2M0107+00	&	L8	&	3	&	15.59	$\pm$	1.1	&	10	&	623	$\pm$	10	&	91	$\pm$	1	&	9	&	$<$	−4.94	&	18	\\
2MASSW J0310599+164816	&	2M0310+16	&	L8	&	1	&	27.1	$\pm$	2.5	&	11	&	245.9	$\pm$	4	&	6.2	$\pm$	3.3	&	11	&	$<$	−5.65	&	18	\\
2MASS J08354256-0819237	&	2M0835-08	&	L6.5	&	4	&	7.21	$\pm$	0.01	&	12	&	-535.657	$\pm$	0.439	&	302.737	$\pm$	0.405	&	12	&	$<$	−7.42 	&	19	\\
2MASS J10101480-0406499	&	2M1010-04	&	L6	&	5	&	16.72	$\pm$	2.27	&	10	&	-321	$\pm$	16	&	20	$\pm$	13	&	14	&	\nodata	&	\nodata	\\
2MASS J10433508+1213149	&	2M1043+12	&	L9	&	3	&	14.6	$\pm$	2.26	&	10	&	26	$\pm$	5.1	&	-234.2	$\pm$	3.9	&	10	&		\nodata	& \nodata	\\
DENIS-P J1058.7-1548	&	DENIS 1058-15 	&	L2.5	&	3	&	18.3	$\pm$	0.18	&	12	&	-258.068	$\pm$	0.809	&	31.104	$\pm$	0.732	&	12	&		−5.59	&	18	\\
2MASS J12195156+3128497	&	2M1219+31	&	L8	&	6	&	18.1 $\pm$ 3.7	& 21	&	-233	$\pm$	23.7	&	-49.6	$\pm$	14.7	&	15	&	\nodata	& \nodata	\\
2MASS J14252798-3650229	&	2M1425-36	&	L5	&	7	&	11.83	$\pm$	0.05	&	12	&	-283.863	$\pm$	0.611	&	-469.283	$\pm$	0.48	&	12	&	$<$	−5.03	&	18	\\
2MASS J16154255+4953211	&	2M1615+49	&	L4$\beta$	&	8	&	31.25	$\pm$	0.98	&	17	&	-80	$\pm$	12	&	18	$\pm$	12	&	16	&	\nodata	& \nodata	\\
2MASS J16322911+1904407	&	2M1632+19	&	L8	&	3	&	15.24	$\pm$	0.49	&	10	&	293	$\pm$	1	&	-54	$\pm$	1	&	13	&	$<$	−5.52	&	18	\\
2MASS J17114573+2232044	&	2M1711+22	&	L9.5	&	3	&	30.2	$\pm$	4.39	&	10	&	31	$\pm$	7	&	-5	$\pm$	4	&	13	&	$<$	−5.39	&	18	\\
2MASSI J1721039+334415	&	2M1721+33 	&	L3	&	9	&	16.31	$\pm$	0.06	&	12	&	-1855.601	$\pm$	0.358	&	591.642	$\pm$	0.369	&	12	&	$<$	−5.51	&	18	\\
2MASS J17502484-0016151	&	2M1750-00	&	L4.5	&	3	&	9.24	$\pm$	0.02	&	12	&	-397.154	$\pm$	0.456	&	197.921	$\pm$	0.402	&	12	&		−6.2 $\pm$ 0.1	&	20	\\
2MASS J18212815+1414010	&	2M1821+14	&	L4.5	&	2	&	9.36	$\pm$	0.02	&	12	&	227.324	$\pm$	0.54	&	-246.409	$\pm$	0.553	&	12	&	\nodata	&	\nodata\\
2MASS J21481628+4003593	&	2M2148+40	&	L6	&	2	&	8.11	$\pm$	0.03	&	12	&	773.298	$\pm$	0.701	&	458.01	$\pm$	0.884	&	12	&	\nodata	&	\nodata
\enddata
\tablerefs{
(1)	\citet{kirkpatrick2000};
(2)	\citet{Metchev2015};
(3)	\citet{schneider2014};
(4)	\citet{salim2003} 
(5)	\citet{cruz2003};
(6)	\citet{Chiu2006};
(7)	\citet{Kendall2007};
(8)	\citet{reid2008};
(9)	\citet{Schmidt2007};
(10)	\citet{Faherty2012};
(11)	\citet{Smart2013};
(12)	\citet{Gaia2018};
(13)	\citet{Faherty2009};
(14)	\citet{Jameson2008};
(15)	\citet{Schmidt2010};
(16)	\citet{Faherty2016};
(17)	\citet{Liu2016};
(18)    \citet{schmidt2015};
(19)    \citet{Reiners2008};
(20)    \citet{Pineda2016};
(21)    \citet{Schmidt2010AJ....139.1808S}.
}
\end{deluxetable*}

\section{Targets}\label{sec:targets}

We selected our sample of 17 objects to include only those with photometric variability at $I, R, J, H,$ and/or $K$ bands to test whether this variability can be attributed to a magnetically driven component in addition to cloudy atmospheres. Table \ref{tab:target_info} presents the target summary. In this work, we chose to focus on L dwarfs; however, in \S \ref{sec:discussion} we discuss combining this work with independent analysis of T dwarf O/IR variability and radio aurorae to yield a correlation over the full range of the L/T spectral sequence. Below we outline the literature in regards to the photometric variability, previous radio searches, and H$\alpha$ activity of each of our targets. We additionally include a summary table as Table \ref{tab:target_var_info}.

\textit{2MASSI J0030300-145033}. 2M0030-14 was discovered and classified as an L6.5 dwarf by \citet{kirkpatrick2000} using data from the Two Micron All Sky Survey (2MASS; \citealt{Skrutskie2006}). \citet{enoch2003} saw a magnitude change of 0.19 $\pm$ 0.11 mag in the $K$ band. Other observations in $izJHK$ bands by \citet{Koenetal2005}, \citet{clarke2008}, \citet{schmidt2015}, \citet{Radigan2014} report no variability. \citet{berger2006} placed an upper limit on its radio emission of 57 $\mu$Jy at 8.46 GHz. \citet{schmidt2015} placed an upper limit on the H$\alpha$ activity of 2M0030-14 of $\text{log}(L_{\text{H}\alpha} / L_{\text{bol}}) < -5.04$.

\textit{2MASSI J0103320+193536}. 2M0103+19 was discovered by \citet{kirkpatrick2000} and is classified as an L6 dwarf \citep{Metchev2015}. \citet{Metchev2015} identified a rotation period of 2.7 $\pm$ 0.1 hr and saw variability in the $Spitzer$ IRAC channels 1 (3.6 $\mu$m) and 2 (4.6 $\mu$m) with magnitude changes of 0.56 $\pm$ 0.03\% and 0.87 $\pm$ 0.09\%, respectively. Additionally, \citet{enoch2003} observed $K$ band variability of 0.10 $\pm$ 0.02 mag and no variability in the $J$ band, the latter of which was confirmed by \citet{vos2019}. \citet{schmidt2015} placed an upper limit on the H$\alpha$ activity of 2M0103+19 of $\text{log}(L_{\text{H}\alpha} / L_{\text{bol}}) < -5.96$.

\textit{2MASS J01075242+0041563}. 2M0107+00 was discovered by \citet{geballe2002} using data from the Sloan Digital Sky Survey (SDSS; \citealt{york2000}) and is classified as an L8 dwarf \citep{schneider2014}. \citet{Metchev2015} observed 2M0107+00 to have an irregular period between 5 -- 13 hr with variability at 3.6 $\mu$m and 4.6 $\mu$m of 1.27 $\pm$0.13\% and 1.0  $\pm$ 0.2\%, respectively. \citet{schmidt2015} placed an upper limit on the H$\alpha$ activity of 2M0107+00 of $\text{log}(L_{\text{H}\alpha} / L_{\text{bol}}) < -4.94$.

\textit{2MASSW J0310599+164816}. 2M0310+16 was discovered by \citet{kirkpatrick2000} who classified it as an L8 dwarf. More recently \citet{stumpf2010} resolved 2M0310+16 as a brown dwarf binary system with a separation of $<$6 AU. Using HST/WFC3 in the IR channel, \citet{buenzli2014} saw an amplitude change of 2\% per hour at 1.26 -- 1.32 $\mu$m. \citet{schmidt2015} placed an upper limit on the H$\alpha$ activity of 2M0310+16 of $\text{log}(L_{\text{H}\alpha} / L_{\text{bol}}) < -5.65$.

\textit{2MASS J08354256-0819237}. 2M0835-08 was identified and classified as an L6.5 dwarf by \citet{salim2003}. It has a known rotation period as seen in the $I$ band of 3.1 hr \citep{koen2004a}. \citet{Radigan2014} reported a 1.3 $\pm$ 0.2\% amplitude variation in the $J$ band, whereas \citet{wilson2014} reported 1.6 $\pm$ 0.5\%. \citet{koen2004a} saw a 10 mmag amplitude in the $I$ band. No variability is seen in the $R$ band \citep{Koenetal2005}. \citet{Schlawin2017} observed 2M0835-08 with SpeX IRTF $JHK$ broad bands from 0.9 -- 2.4 $\mu$m and placed an upper limit of $<$ 0.5\% semi-amplitude in each band. \citet{berger2006} reported a non-detection searching for radio emission with a sensitivity of 30 $\mu$Jy. \citet{Reiners2008} placed an upper limit on the H$\alpha$ activity of 2M0835-08 to be $\text{log}(L_{\text{H}\alpha} / L_{\text{bol}}) < -7.42$. More recently, \citet{schmidt2015} placed the upper limit on the H$\alpha$ activity at -6.60, which is similar to the upper limit of -6.5 seen by \citet{Pineda2016}.

\textit{2MASS J10101480-0406499}. 2M1010-04 was discovered by and identified as an L6 dwarf by \citet{cruz2003}. \citet{wilson2014} reported the variability in the $J$ band to be 5.1 $\pm $1.1\%; however, the data was reanalyzed independently by \citet{Radigan2014} who found it to be 3.6 $\pm$ 0.4\%. There have been no H$\alpha$ observations of 2M1010-04.

\textit{2MASS J10433508+1213149}. 2M1043+12 was discovered by \citet{Chiu2006} using SDSS data and classified as an L9 dwarf by \citet{schneider2014}. \citet{Metchev2015} determined an irregular rotation period of 3.8 $\pm$ 0.2 hr with a variation in the $Spitzer$ IRAC channels 1 and 2 of 1.54 $\pm$ 0.15\% and 1.2 $\pm$ 0.2\%, respectively. There have been no H$\alpha$ observations of 2M1043+12.

\textit{DENIS-P J1058.7-1548}. DENIS 1058-15 was discovered by \citet{Tinney1997} and was classified as an L2.5 dwarf by \citet{schneider2014}. \citet{Heinze2013} report a rotation period of $4.25^{+0.26}_{-0.16}$ hr with a variability amplitude of 0.39 $\pm$ 0.04\% at 3.6 $\mu$m and 0.090 $\pm$ 0.056\% at 4.5 $\mu$m. The authors also determine an amplitude of 0.843 $\pm$ 0.098\% in the $J$ band with a rotation period of 4.31 hr. \citet{Metchev2015} independently confirmed the IRAC amplitudes, measuring a rotation period of 4.1 $\pm$ 0.2 hr. Observations by \citet{Koen2013} reveal no variability in the $IR$ bands. \citet{schmidt2015} measured the H$\alpha$ activity of DENIS 1058-15 to be $\text{log}(L_{\text{H}\alpha} / L_{\text{bol}}) = -5.59$.

\textit{2MASS J12195156+3128497}. 2M1219+31 was identified and classified as an L8 dwarf by \citet{Chiu2006}. There is currently no measured rotation period. \citet{buenzli2014} tentatively report a $\sim$3 -- 6\% per hour amplitude variation from 1.12 -- 1.20 $\mu$m and no variability from 1.32 -- 1.66 $\mu$m. There have been no H$\alpha$ observations of 2M1219+31.

\begin{deluxetable*}{l c c c c c c c c c c c c}[t]
\centering
\setlength{\tabcolsep}{0.06in}
\tabletypesize{\scriptsize}
\tablecaption{\normalsize{Observation summary of our sample.}\label{tab:obs_sum}} 
\tablehead{
\colhead{} \vspace{-0.3cm} & 
\colhead{} & 
\colhead{Obs.} & 
\colhead{Obs.} & 
\colhead{Time} & 
\colhead{VLA} & 
\colhead{Synthesized} & 
\colhead{$f_{\nu}$ \tablenotemark{a}} & 
\colhead{$f_{\nu}$ \tablenotemark{a}} & 
\colhead{$\log_{10}(L_{\nu})$}  & 
\colhead{Phase} & 
\colhead{Flux}\\
%
% \vspace{-0.3cm}
\colhead{Object} \vspace{-0.3cm} & 
\colhead{Band} & 
\colhead{Date} & 
\colhead{Block} & 
\colhead{on} & 
\colhead{Config.} & 
\colhead{Beam Size} & 
\colhead{Stokes $I$} & 
\colhead{Stokes $V$} & 
\colhead{Stokes $I$, $V$} & 
\colhead{Calibrator} & 
\colhead{Calibrator}\\
%
% \vspace{-0.3cm}
\colhead{} \vspace{-0.3cm} & 
\colhead{[GHz]} & 
\colhead{(2016)} & 
\colhead{[h]} & 
\colhead{[s]} & 
\colhead{} & 
\colhead{[$\arcsec$ $\times$ $\arcsec$]} &
\colhead{[$\mu$Jy]} & 
\colhead{[$\mu$Jy]} & 
\colhead{[erg s$^{-1}$ Hz$^{-1}$]} & 
\colhead{} & 
\colhead{} \\ \vspace{-0.2cm}
}
\startdata
2M0030-14		&		4--8		&		1-May		&	2	&	5470	&	CnB		            &	2.78 $\times$ 1.54    &	$<$	17.4	&	$<$	11.4	    &   $<$ 13.2, $<$ 13.0    &   J0050-0929		&	3C48	\\
2M0103+19		&		4--8		&		8-Apr		&	2	&	5644	&	C		            &	2.98 $\times$ 2.77	&	$<$	11.4	&	$<$	9.9	        &   $<$ 12.8, $<$ 12.7    &   J0112+2244		&	3C48	\\
2M0107+00		&		4--8		&		2-Jun		&	2	&	5290	&	B		            &	1.01 $\times$ 0.92	&	$<$	10.2	&	$<$	9.0	        &   $<$ 12.5, $<$ 12.4    &   J0059+0006		&	3C48	\\
2M0310+16		&		4--8		&		6-May		&	2	&	5546	&	CnB		            &	3.41 $\times$ 0.99	&	$<$	10.8	&	$<$	10.2    	&   $<$ 13.0, $<$ 13.0    &   J0318+1628		&	3C48	\\
2M0835-08		&		4--8		&		6-Jun		&	2	&	5112	&	B		            &	1.59 $\times$ 0.93	&	$<$	14.7	&	$<$	13.8	    &   $<$ 12.0, $<$ 11.9    &   J0820-1258		&	3C286	\\
2M1010-04		&		4--8		&		10-May		&	2	&	5408	&	CnB		            &	3.21 $\times$ 1.29	&	$<$	47.4	&	$<$	14.4	    &   $<$ 13.2, $<$ 12.7    &   J1024-0052		&	3C286	\\
2M1043+12		&		4--8		&		14-Jun		&	2	&	4992	&	B		            &	0.95 $\times$ 0.88	&	$<$	12.6	&	$<$	12.0	    &   $<$ 12.5, $<$ 12.5    &   J1120+1420		&	3C286	\\
DENIS 1058-15 	&		4--8		&		10-May		&	2	&	5466	&	CnB		            &	3.47 $\times$ 1.45	&	$<$	10.5	&	$<$	9.9	        &   $<$ 12.6, $<$ 12.6    &   J1039-1541		&	3C286	\\
2M1219+31		&		4--8		&		11-Jun		&	2	&	5290	&	B		            &	1.18 $\times$ 0.88	&	$<$	14.1	&	$<$	13.2	    &	$<$ 12.7, $<$ 12.7    &   J1221+2813		&	3C286	\\
2M1425-36		&		4--8		&		5-May		&	2	&	5524	&	CnB		            &	3.13 $\times$ 2.46	&	$<$	12.9	&	$<$	12.0	    &	$<$ 12.3, $<$ 12.3    &   J1356-3421		&	3C147	\\
2M1615+49		&		4--8		&		4-May		&	2	&	5670	&	CnB		            &	2.98 $\times$ 1.01	&	$<$	9.0	    &	$<$	9.6	        &	$<$ 13.0, $<$ 13.0    &   J1620+4901		&	3C147	\\
2M1632+19		&		4--8		&		15-May		&	2	&	5406	&	CnB		            &	3.71 $\times$ 1.06	&	$<$	10.8	&	$<$	11.1	    &	$<$ 12.5, $<$ 12.5    &   J1640+1220		&	3C286	\\
2M1711+22		&		4--8		&		17-May		&	2	&	5406	&	CnB $\rightarrow$ B	&	1.52 $\times$ 0.86	&	$<$	11.4	&	$<$	9.6	        &	$<$ 13.1, $<$ 13.0    &   J1716+2152		&	3C286	\\
2M1721+33 		&		4--8		&		4-May		&	2	&	5520	&	CnB		            &	5.62 $\times$ 1.03	&	$<$	11.4	&	$<$	11.1	    &	$<$ 12.6, $<$ 12.5    &   J1721+3542		&	3C286	\\
2M1750-00		&		4--8		&		5-May		&	2	&	5424	&	CnB		            &	3.5 $\times$ 1.19	    & 185 $\pm$ 18	&	-88 $\pm$ 11	&	13.3,     13.0 	      &   J1804+0101		&	3C286	\\
2M1821+14		&		4--8		&		14-Aug		&	2	&	5262	&	B		            &	1.67 $\times$ 0.86	&	$<$	12.9	&	$<$	12.3	    &	$<$ 12.1, $<$ 12.1    &   J1824+1044		&	3C286	\\
2M2148+40		&		4--8		&		30-May		&	2	&	5230	&	B		            &	0.92 $\times$ 0.77	&	$<$	9.6	&	$<$	10.2        	&	$<$ 11.9, $<$ 11.9    &   J2202+4216		&	3C48	\\
\enddata
\tablenotetext{a}{Upper limits are 3$\sigma_{\mathrm{rms}}$ where $\sigma_{\mathrm{rms}}$ is the rms noise in each image.  For measured flux densities, positive and negative values correspond to right and left circular polarization, respectively. }
\end{deluxetable*}

%check phase calibrator of 0835

\textit{2MASS J14252798-3650229}. 2M1425-36 was discovered by \citet{Kendall2004} and is an L3 dwarf in the optical \citep{Siegler2007} and an L5 dwarf in the IR \citep{Kendall2007}. \citet{Radigan2014} measure a rotation period of 3.7 $\pm$ 0.8 hr based on $J$ band variability with an amplitude of 0.6 $\pm$ 0.1\%. \citet{vos2019} similarly report a $J$ band variability amplitude of 0.7 $\pm$ 0.3. \citet{schmidt2015} placed an upper limit on the H$\alpha$ activity of 2M1425-36 of $\text{log}(L_{\text{H}\alpha} / L_{\text{bol}}) < -5.03$.

\textit{2MASS J16154255+4953211}. 2M1615+49 was discovered by and classified as an L4$\beta$ by \citet{reid2008}. Using the $Spitzer$ IRAC channel 1 and channel 2, \citet{Metchev2015} identify photometric amplitudes of 0.9 $\pm$ 0.2\% and $<$0.39\% in these channels, respectively. The authors also report a rotation period of $\sim$24 hours. \citet{vos2019} observe no variability in the $J$ band. There have been no H$\alpha$ observations of 2M1615+49.

\textit{2MASS J16322911+1904407}. 2M1632+19 was discovered by \citet{kirkpatrick1999} and is an L8 dwarf \citep{schneider2014}. While no variability has been reported in the $JH$ bands \citep{buenzli2014, wilson2014}, \citet{Metchev2015} observed variability amplitudes of 0.42 $\pm$ 0.08\% at 3.6 $\mu$m and 0.5 $\pm$ 0.3\% at 4.5 $\mu$m. The authors determined a regular rotation period of 3.9 $\pm$ 0.2 hr.  Two previous surveys have searched for auroral emission but were only able to report upper limits: \citet{Route2013} observed at 5 GHz with Arecibo and placed a limit of $<$54 $\mu$Jy, while \citet{Antonova2008} used the VLA at 4.9 GHz to place a limit of $<$39 $\mu$Jy. \citet{schmidt2015} placed an upper limit on the H$\alpha$ activity of 2M1632+19 of $\text{log}(L_{\text{H}\alpha} / L_{\text{bol}}) < -5.52$.

\textit{2MASS J17114573+2232044}. 2M1711+22 was discovered by \citet{kirkpatrick2000} and was identified as an L9.5 dwarf by \citet{schneider2014}. \citet{Khandrika2013} report $JK$ variability at 0.103 $\pm$ 0.041 mag semi-amplitude and 0.593 $\pm$ 0.083 mag semi-amplitude, respectively. \citet{buenzli2014} see no variability in the $J$ band. \citet{schmidt2015} placed an upper limit on the H$\alpha$ activity of 2M1711+22 of $\text{log}(L_{\text{H}\alpha} / L_{\text{bol}}) < -5.39$.

\textit{2MASSI J1721039+334415}. 2M1721+33 was discovered by \citet{cruz2003} and is an L3 dwarf \citep{Schmidt2007} with a rotation period of 2.6 $\pm$ 0.1 hr \citep{Metchev2015}. \citet{Metchev2015} observed amplitude variations of 0.33 $\pm$ 0.07\% at 3.6 $\mu$m and $<$0.29\% at 4.5 $\mu$m. \citet{berger2006} searched for radio activity and report an upper limit of 48 $\mu$Jy. \citet{schmidt2015} placed an upper limit on the H$\alpha$ activity of 2M1721+33 of $\text{log}(L_{\text{H}\alpha} / L_{\text{bol}}) < -5.51$.

\textit{2MASS J17502484-0016151}. 2M1750-00 was discovered by \citet{Kendall2007} and classified as an L4.5 dwarf by \citet{schneider2014}. Its rotation period is currently unknown, but  \citet{buenzli2014} observed a photometric amplitude change of $\sim$0.7\% per hour using HST/WFC3 in the $J$ broad band over a 40 min observation period. They report no variability in the $H$ broad band. \citet{Koen2013} and \citet{Radigan2014} observed no variability in the $IR$ and $J$ bands, respectively. \citet{antonova2013} searched for radio emission using the VLA but report a non-detection with 43 $\mu$Jy sensitivity. Additionally, \citet{Pineda2016} measured the H$\alpha$ activity of 2M1750-00 to be $\text{log}(L_{\text{H}\alpha} / L_{\text{bol}}) = -6.2 \pm 0.1$.

\textit{2MASS J18212815+1414010}. 2M1821+14 was discovered by \citet{Looper2008} and was classified as an L4.5 dwarf \citep{Metchev2015}. \citet{Metchev2015} determine an irregular rotation period of 4.2 $\pm$ 0.1 hr, with photometric amplitudes of 0.54 $\pm$ 0.05\% at 3.6 $\mu$m and 0.71 $\pm$ 0.14\% at 4.5 $\mu$m. \citet{yang2015} observed from 1.1 -- 1.7 $\mu$m, seeing a 1.77 $\pm$ 0.11\% amplitude out of the water band (1.4 $\mu$m), and a 1.54 $\pm$ 0.21\% amplitude at the water band. In a dedicated study, \citet{Schlawin2017} demonstrated that there is a steady decrease in variability amplitude from 0.9 -- 2.4 $\mu$m starting at 1.5\% semi-amplitude at 0.9 $\mu$m, eventually decreasing to 0\% at $\sim$1.7 $\mu$m, where it remains through 2.4 $\mu$m. \citet{Koen2013} reported no variability in the $IR$ bands. There have been no H$\alpha$ observations of 2M1821+14.

\textit{2MASS J21481628+4003593}. 2M2148+40 was discovered by \citet{Looper2008} and was classified as an L6 dwarf \citep{Metchev2015}. \citet{Metchev2015} determine a rotation period of 19 $\pm$ 4 hr, with photometric amplitudes of 1.33 $\pm$ 0.07\% at 3.6 $\mu$m and 1.03 $\pm$ 0.1\% at 4.5 $\mu$m. \citet{Khandrika2013} report no variability in the $J$ band. \citet{antonova2013} report a non-detection at 4.9 GHz using the VLA with a sensitivity of 63 $\mu$Jy. There have been no H$\alpha$ observations of 2M2148+40.

\begin{figure*}[t]
\centering
\plottwo{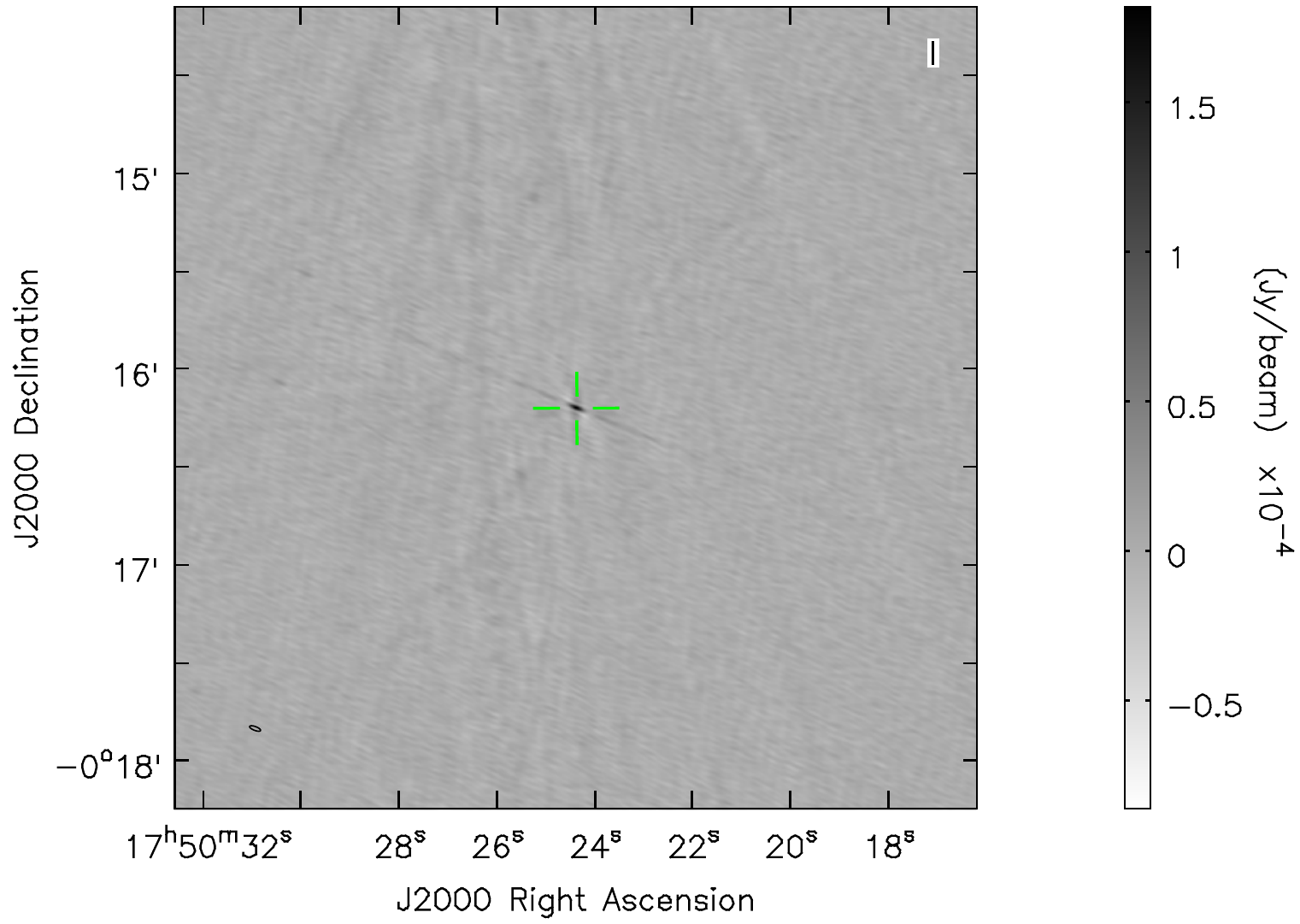}{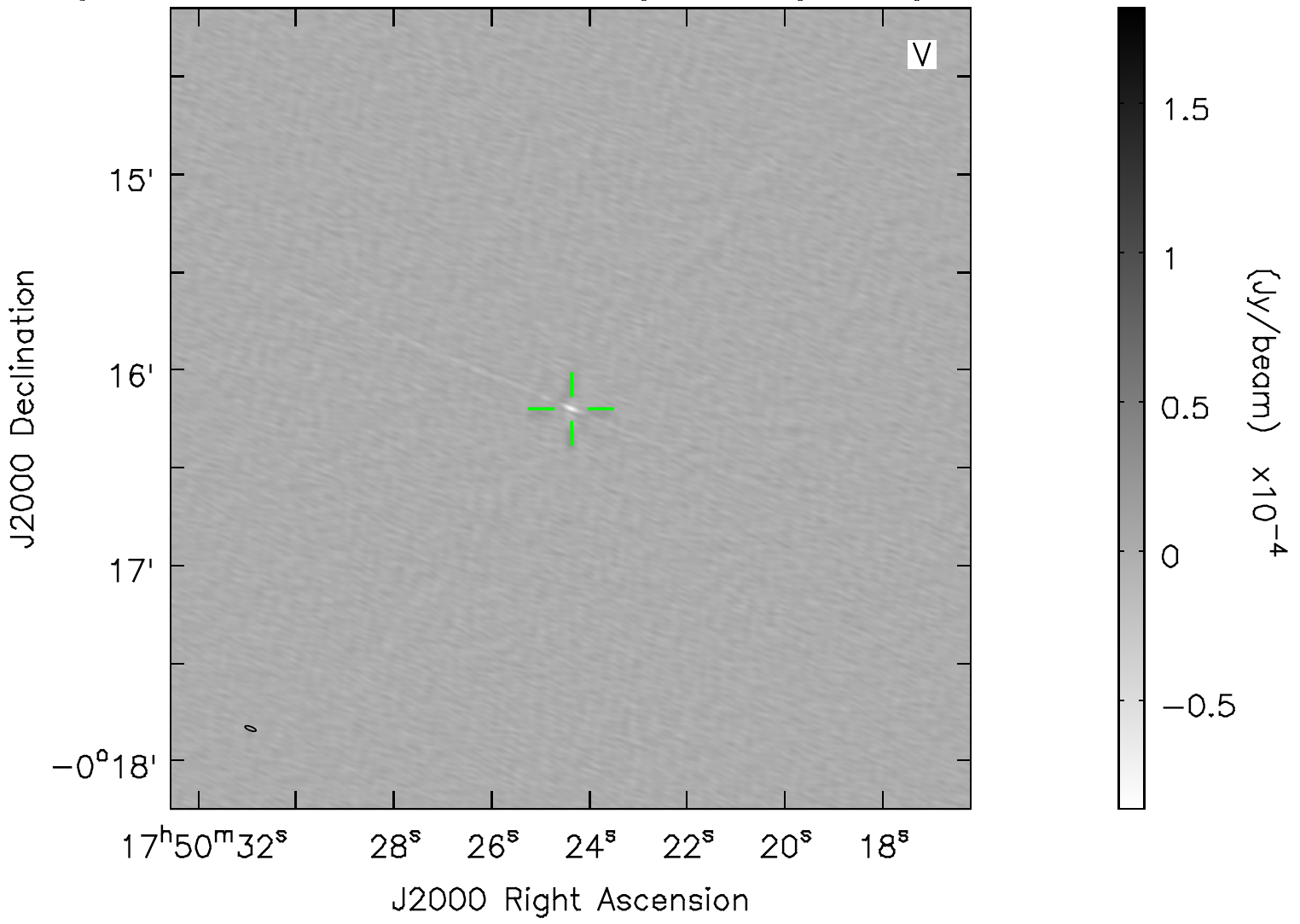}
\caption{Stokes $I$ (left) and Stokes $V$ (right) images of 2M1750-00. The cross-hairs denote the calculated proper-motion-corrected coordinates of our target. The synthesized beam is seen in the lower-left corner.}
\label{fig:J1750-IV}
\end{figure*}

\section{Observations}\label{sec:observations}

We observed the 17 targets with the Karl G. Jansky Very Large Array (VLA) at C-band (4 -- 8 GHz). We used the WIDAR correlator in 3-bit observing mode for 4 GHz bandwidth observations with 2 s integrations in 2-hour time blocks for 34 total program hours. We used the full 4 GHz bandwidth available to achieve $\sim$3 $\mu$Jy sensitivity. Observations were made between April -- August 2016 at C, CnB (i.e. while the VLA was moving from C to B configuration), and B configurations. Since our targets are point sources and not resolved, the configuration did not affect the results of our survey.  The observations are summarized in Table \ref{tab:obs_sum}. 

\subsection{Calibrations}

We calibrated our measurement sets using \texttt{Common Astronomy Software Applications (CASA)} version 5.6.1-8 packages. Raw measurement sets were calibrated with the VLA Calibration Pipeline  using the flux and phase calibrators in Table \ref{tab:obs_sum}, after which we manually flagged the radio frequency interference (RFI). Flux calibrators were observed once during each observing block, and flux bootstrapping results in an absolute flux calibration accuracy of $\sim$5\%. Phase calibrators were within 10 degrees of each target and of S or P quality for C-band at our configurations. To calibrate complex gain solutions, we alternated between phase calibrator and target with cycle times of $\sim$30 min.

\subsection{Source Motion}

We corrected the 2MASS coordinates \citep{Skrutskie2006} of our targets to determine expected positions using the proper motion measurements listed in Table \ref{tab:obs_sum}. We phase-centered each object to these coordinates with \texttt{fixvis} before using the \texttt{clean} routine to image each target.

\section{Results}\label{sec:results}

\subsection{Imaging}

\begin{figure*}[t!]
\centering
\plottwo{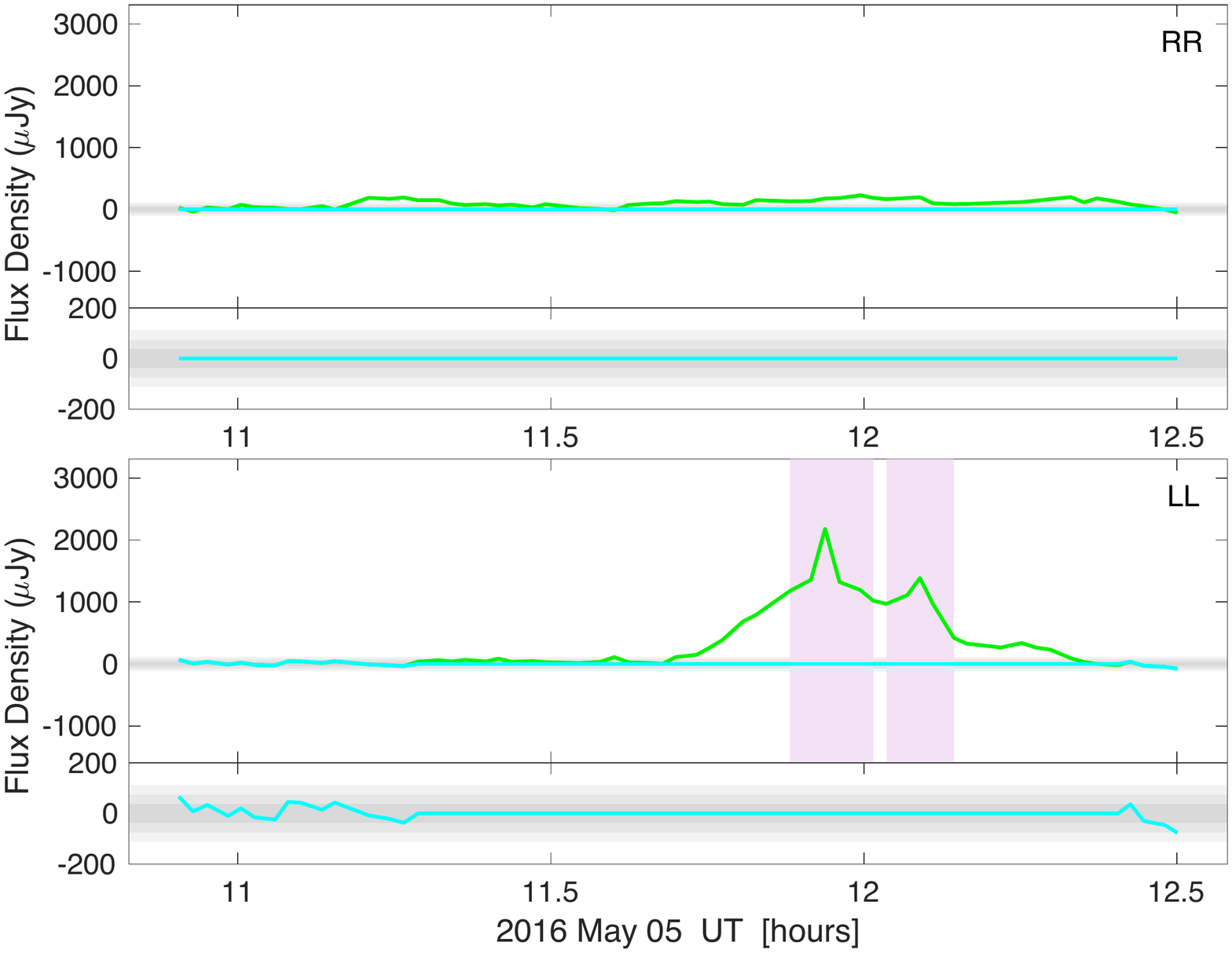}{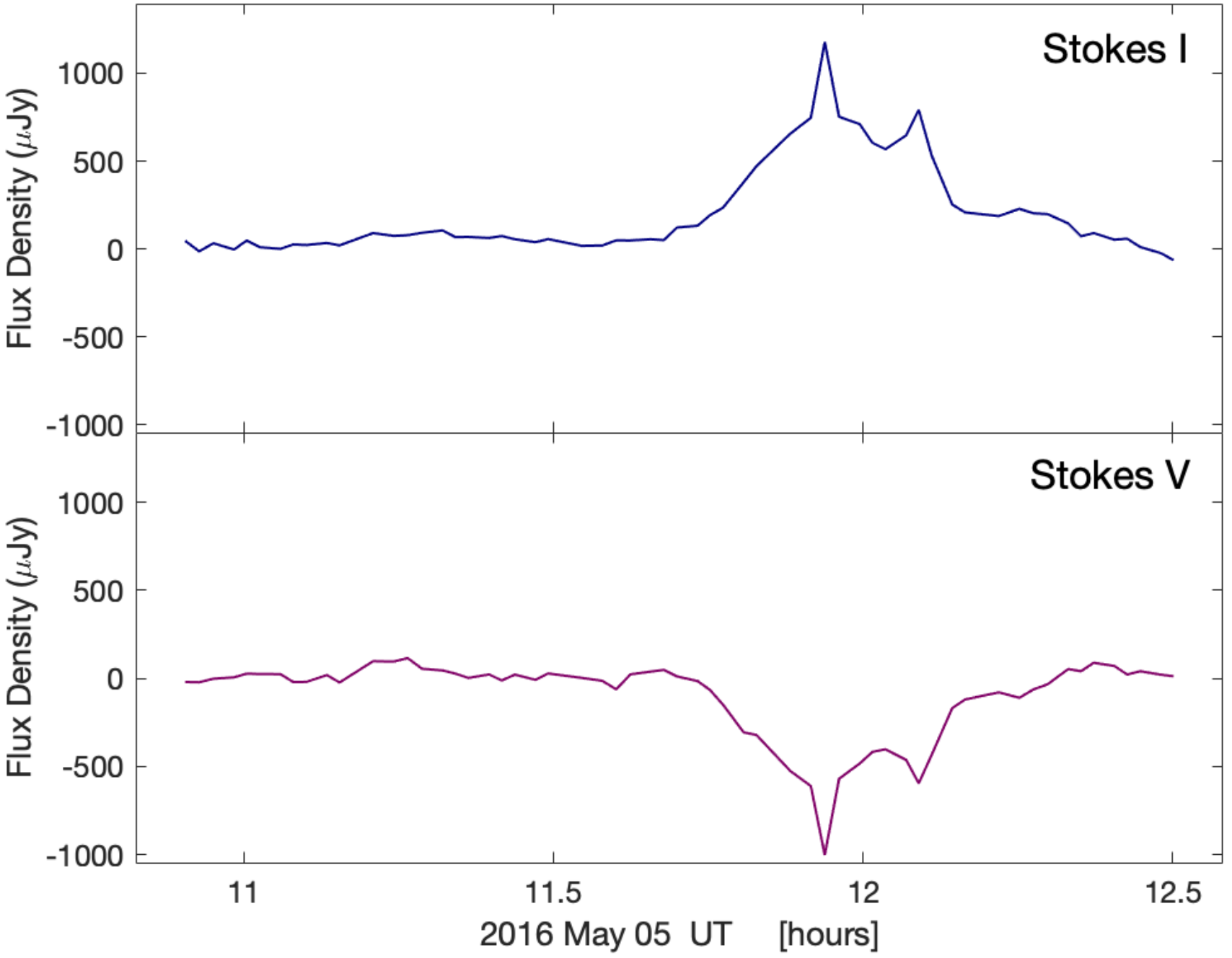}
\caption{(Left) The right- and left-handed correlations of 2M1750-00 from 4 -- 8 GHz with 2 second integrations. The green lines represent the smoothed data while the cyan line shows the level of quiescent emission after removing the circularly polarized flare. Gray regions are the 1$\sigma$, 2$\sigma$, and 3$\sigma$ detection limits. We see one pulse over the course of the two-hour observation, implying a rotation period $\geq$ 2 hours. There is a clear broad peak in emission that has definite sub-structure (highlighted). (Right) Same as left image but presented in Stokes $I$ and Stokes $V$, total intensity and circularly polarized emission, respectively.}
\label{fig:2M1750-IV}
\end{figure*}

We produced Stokes $I$ and Stokes $V$ (total and circular polarization, respectively) images using the \texttt{CASA} task \texttt{clean}. We used a Brigg's weighting of 0.0, which gives a good trade-off between sensitivity and resolution. We searched for a point source at the proper-motion-corrected coordinates in each image. We self-calibrated one target, 2M1010+00, to mitigate phase errors in  three brighter ($\sim$3 -- 11 mJy) objects in the field and improved the rms noise in the image by 25\%. We detect radio emission in Stokes $I$ and Stokes $V$ from one target in our sample, 2M1750-00.

\begin{deluxetable}{l c c c}[h]
\centering
\tabletypesize{\footnotesize}
\tablecaption{\normalsize{Time- and frequency-integrated flux density measurements of 2M1750-00.}\label{tab:results}} 
\tablehead{  \colhead{Temporal} \vspace{-0.3cm} & \colhead{$f_{\nu}$, Stokes $I$} & \colhead{$f_{\nu}$, Stokes $V$} & \colhead{Circ. Pol.} \\
\colhead{Segment} & \colhead{[$\mu$Jy]} & \colhead{[$\mu$Jy]} & \colhead{[\%]}
}
\startdata
All &	185 $\pm$ 18 &	-88 $\pm$ 11 &	-47.1$_{-30.0}^{+18.9}$	\\
Peak 1 &	926 $\pm$ 40 &	-667 $\pm$ 26 &	-72.0$_{-14.3}^{+10.3}$	\\
Peak 2 &	487 $\pm$ 20 &	-355 $\pm$ 27 &	-72.8$_{-20.4}^{+17.7}$	\\
Quiescent &	56.4 $\pm$ 5.5 &	31.9 $\pm$ 6.6 &	56.0$_{-35.0}^{+41.5}$ \\
\enddata
\end{deluxetable}

We used the \texttt{CASA} task \texttt{imfit} to determine the position for 2M1750-00 and measure the mean flux density by fitting an elliptical Gaussian point source to the cleaned image. The mean flux density was 185 $\pm$ 18 $\mu$Jy (S/N $\sim$40) in Stokes $I$ and -88 $\pm$ 11 $\mu$Jy (S/N $\sim$25) in Stokes $V$ and is unresolved, with a source size of $3\farcs61  \times 1\farcs14$. $3\sigma$ upper limits for undetected sources are listed in Table \ref{tab:obs_sum}.

\begin{figure*}[t]
    \centering
    \includegraphics[width=0.75\linewidth]{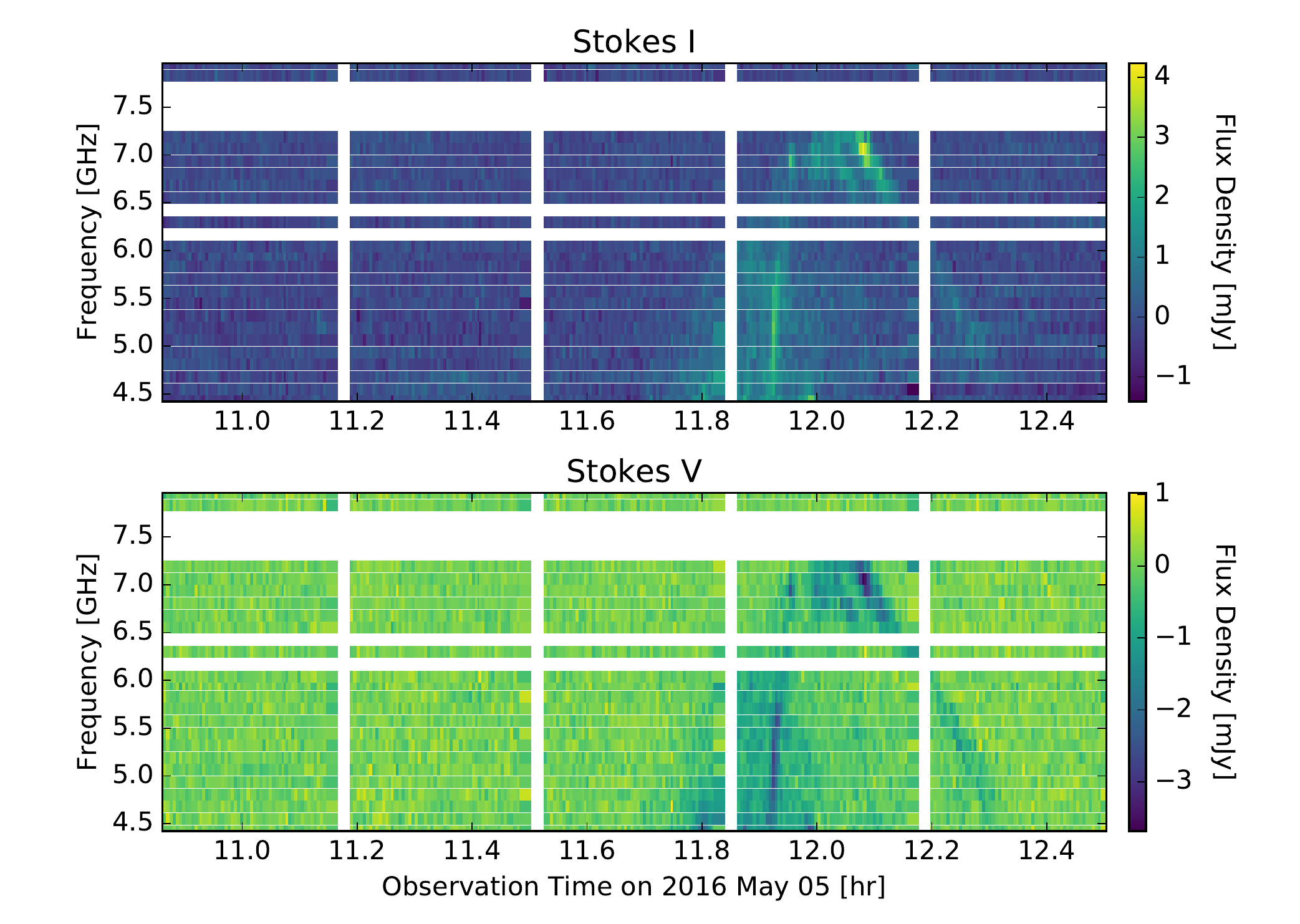}
    \caption{The dynamic spectrum of 2M1750-00 with flux density relative to the average. White regions represent the times where phase calibrator observations took place or frequencies at which significant data was removed due to RFI. We find that the ECM flare from this object is characterized by a broad peak in emission that has sub-structure. The emission occurs throughout the entire 4 -- 8 GHz bandwidth, implying a magnetic field strength $\geq$2.9 kG.}
    \label{fig:dynamic_spectrum}
\end{figure*}

\subsection{Time Series}

In addition to visual inspection, we performed a time series analysis of each target. Following the procedure outlined in \citet{kao2018}, we used the \texttt{CASA} task \texttt{plotms} to export the real uv visibilities in the rr and ll correlations, averaged across all baselines, channels, and spectral windows.
%, because the S/N can be up to $\sqrt{2}$ higher in each of the individual correlations.
We created time series of both the rr and ll correlations and calculated the Stokes $I$ and $V$ flux densities as a function of time averaged over the entire 4 -- 8 GHz bandwidth. 

We additionally averaged the measurement sets with time resolutions of 10 s, 30 s, and 60 s and frequency resolutions of 2 GHz to search for emission that may have been averaged out in the time-averaged images. We do not detect any statistically significant radio emission in the time series of any additional objects. For 2M1750-00, timeseries at all time resolutions show a single, highly circularly polarized flare with a double-peaked morphology (Figure \ref{fig:2M1750-IV}), implying a rotation period $\geq 2$ hours.

For each peak in the observed flare from 2M1750-00, we image over the full-width half-max (FHWM) of the peak and measure the average Stokes $I$ and Stokes $V$ flux densities of the flaring emission using the \texttt{CASA} task \texttt{imfit}.  We additionally measure the non-flaring quiescent emission by subtracting the full width of the peak, defined as three times the FWHM of each peak of the flare, from the data. 

We report the polarization characteristics of the flaring and non-flaring emission in Table \ref{tab:results}.  The flux densities of the peaks of the flare are between 8 -- 17 times stronger than quiescent emission in Stokes $I$ and 11 -- 22 times stronger in Stokes $V$. The fractional circular polarization for the flaring emission
is 72.0\%$_{-14.3}^{+10.3}$ for the first peak and 72.8\%$_{-20.4}^{+17.7}$ for the second peak, consistent with measurements of highly circularly polarized ECM emission seen by \citet{hallinan2007} and theoretically predicted by \citet{Treumann2006}.

\subsection{Dynamic Spectrum of 2M1750-00}
We explore the frequency and temporal dependencies of the flare from 2M1750-00 by creating a dynamic spectrum (Figure \ref{fig:dynamic_spectrum}). Using the \texttt{CASA} task \texttt{plotms}, we exported the real uv visibilities in the rr and ll correlations, averaged across all baselines and channels. We then calculated the Stokes $I$ and $V$ flux densities for each time and frequency element. Four main vertical gaps in time are marked in white where the phase calibrator observations took place, along with three horizontal gaps in frequency where a significant amount of data was flagged and removed due to RFI. We find that there is clear substructure within the one flare and that emission continues throughout the entire 4 -- 8 GHz bandwidth. 
Similar cases of substructure have been observed in the radio aurorae for LSR J1835+3259 \citep{hallinan2015} and 2M1047+21 \citep{Williams2015, kao2018}. The underlying mechanism of this substructure remains unknown; however, \citet{hallinan2015} speculate that this substructure is likely due to contributions from individual, small-scale current systems, similar to what was surmised for the fine structure in the auroral kilometric radiation observed from Jupiter and Saturn \citep[][and references therein]{Gurnett1981, Pottelette1999, Treumann2006}. Given the prevalence of such substructure, understanding the physical driving mechanism of this emission is imperative.

\subsection{Lower limits on the large-scale magnetic field strength of 2M1750-00}

The disk-averaged brightness temperature of the detected flare from 2M1750-00 is $\geq10^{12.6}$ K. Full rotational phase coverage is needed to confirm if the observed flare is periodic on rotational timescales.  Nevertheless, the short duration of the flare compared to its $\gtrsim 160$ min rotation period, inferred from infrared photometric monitoring \citep{buenzli2014}, is consistent with a flare source region that is much smaller than the disk size of the dwarf.  This high minimum brightness temperature together with the strong circular polarization observed during the pulse is consistent with a coherent  emission process, as is the case for plasma or ECMI emission.  

Plasma emission is emitted at the local plasma frequency $\nu_{\mathrm{pe}} = [e^2 n_e / (\pi m_e)]^{1/2} \approx 9n_e^{1/2}$ kHz or its second harmonic 2$\nu_{\mathrm{pe}}$.  The 4 -- 8 GHz flare detected from 2M175-00 would imply coronal plasma densities on the order of $n_e \sim 10^{11}$ for plasma emission, which exceed expected densities for active M dwarfs \citep{Villadsen2019}. Recent detections of white-light flares on an L2.5 dwarf \citep{Jackman2019} demonstrate that strong flares persist in early L dwarfs.  However, such flares occur less frequently in early L dwarfs compared to M dwarfs \citep[e.g.][]{Schmidt2016, Paudel2018, Paudel2020MNRAS.494.5751P}.  Furthermore, X-ray emission that correlates with hot coronal plasma is underluminous for L dwarfs compared to their radio emission \citep{williams2014}.   This suggests that the plasma densities in L dwarf atmospheres are less than those of active M dwarfs, which can emit electron cyclotron maser emission \citep{OstenBastian2006, Villadsen2019}.  We conclude that the flare observed on 2M1750-00 is likely attributable to ECMI emission.   

For low plasma densities where the ratio of plasma frequency to cyclotron frequency $< 0.3$, 
ECM instability emission is expected to be produced at the fundamental cyclotron frequency $\nu_{\text{[MHz]}}$ \citep{Melrose1984, Treumann2006}:

\begin{equation}
    \nu_{\text{[MHz]}} \sim 2.8 \times B_{\text{[Gauss]}}.
\end{equation}

\noindent The flare on 2M1750-00 persists throughout our frequency band between 4 -- 8 GHz.   If the observed flare from 2M1750-00 is indeed produced via the electron cyclotron maser instability, we can constrain the local magnetic field strength of the brown dwarf to $\geq$ 2.9 kG. Observations of 2M1750-00 above 8 GHz will be required to assess the upper limit of the magnetic field strength of this target.

\subsection{Occurrence Rates of Quiescent Radio Activity}

While the detection rate of our survey agreed with typical volume-limited surveys at $\sim$6\%, we also calculate the underlying occurrence rate of quiescent radio emission. Detectable levels of quiescent radio emission have been observed in all previous observations of periodically pulsed auroral emission, and is therefore considered a proxy for auroral activity. While the source of the quiescent emission is unconfirmed, it has been speculated that it may trace extrasolar analogs to the Jovian radiation belts, where high-energy electrons are trapped by the magnetosphere \citep{hallinan2006, pineda2017, kao2019}. The large (kG) magnetic fields of brown dwarfs and surrounding plasma radiation belts may provide the necessary powerhouse and electron reservoir for both the quiescent emission and auroral ECMI emission \citep{Pineda2018H, kao2019}. 

With this aim, we utilized the maximum likelihood occurrence rate calculation framework developed by \citet{kao2020}. This generalized calculation takes into account each object's distance, observational sensitivity, and an assumed intrinsic radio luminosity distribution.   For the latter, we assume a uniform distribution over previously observed ultracool dwarf (M7 or later spectral type) quiescent radio luminosities. Detected L and M dwarf luminosities overlap in luminosity range, with $[L_{\nu}]$ between 12.6--13.6 and 12.4--13.6 erg s$^{-1}$ Hz$^{-1}$, respectively \citep{kao2020}. In contrast, detected T dwarf luminosities have so far been fainter than detected L dwarf luminosities, with  $[L_{\nu}] \in [11.7,12.7]$ erg s$^{-1}$ Hz$^{-1}$.  Assuming a uniform distribution over the full $[L_{\nu}] \in [11.7,13.6]$ erg s$^{-1}$ Hz$^{-1}$ luminosity range for detected ultracool dwarf radio emission accounts for the possibility of fainter and heretofore undetected L dwarf emission.

Following \citet{kao2020},  we assume a minimum signal-to-noise ratio of 4 for confirmed radio detections and compute the probability density distributions of quiescent radio emission occurrence rates between $[0, 1]$ for each given sample of brown dwarfs.  Simulations of sample sizes with 10 and 20 objects show that on average the  quiescent radio occurrence rate formalism recovers the simulated emission rate of quiescent radio emission in the population, better than does a detection rate. This is especially the case for samples with rms sensitivites that are on average lower than the literature distribution, which is the case when we include our presented observations. 

The two main samples that we compare are L dwarfs with low- and high-amplitude variability. Dissimilar distributions would suggest that high-amplitude variability may be a viable tracer of quiescent radio emission. In the absence of existing empirical measurements of the relative increase in photometric variability amplitudes that may be caused by  energy deposition from magnetic field-aligned currents, we test  variability amplitude cutoffs between 1--3\%.

Table \ref{tab:var_info} shows our input sample of 77 L dwarfs that have been observed at radio frequencies and adhere to the data inclusion policy outlined in \citet{kao2020}. The majority (48) of these objects have been observed for O/IR variability, for which the amplitude, wavelength, and periodicity information is listed. Note that some objects have multiple observations in the same bandpass with both a detection and non-detection of O/IR variability. However, we do not expect the stability of O/IR variability to significantly impact the presence of quiescent radio emission, for which observations confirm can persist for at least 10 years \citep[e.g.][]{hallinan2006, Gawronski2017}, since the underlying driving mechanisms are different.
%though we note two exceptions in the existing literature where  the L2.5 dwarf 2MASS J05233822-1403022 \citep{antonova2007, berger2006, berger2010} and the M9.5 dwarf BRI~0021 \citep{berger2010} show long term variability in the quiescent emission at C-band frequencies.}
In cases where data was re-examined, we defer to the updated results. Finally, we remove all binary objects, as binaries may demonstrate a different occurrence rate distribution than that of single objects \citep{kao2020}. The number of targets in each sample for each cutoff is seen in Table \ref{tab:numbers}.

\begin{deluxetable}{l c c}[h]
\setlength{\tabcolsep}{0.2in}
\centering
\tabletypesize{\normalsize{}}
\tablecaption{Number of objects used in each sample with varying photometric amplitude cutoff. \label{tab:numbers}} 
\tablehead{
% \vspace{-0.3cm}
\colhead{ } & \multicolumn{2}{l}{Number in Sample} \\
\vspace{-0.2cm} \\
\colhead{Amp.} & \colhead{No/Low-} & \colhead{High-} \\
\colhead{Cutoff} & \colhead{Amp.\tablenotemark{a}} & \colhead{Amp.\tablenotemark{b}}
}
\startdata
1\%	&	23	&	12	\\
1.5\%	&	26	&	9	\\
2\%	&	28	&	7	\\
2.5\%	&	28	&	7	\\
3\%	&	30	&	5	\\
\enddata
\tablenotetext{a}{No or Low-Amplitude is defined as variability at the percentage below the amplitude cutoff.}
\tablenotetext{b}{High-Amplitude is defined as variability at the percentage above the amplitude cutoff.}
\end{deluxetable}
%\vspace{-0.3cm}

\begin{figure*}[t]
    \centering
        \includegraphics[width=.9\linewidth]{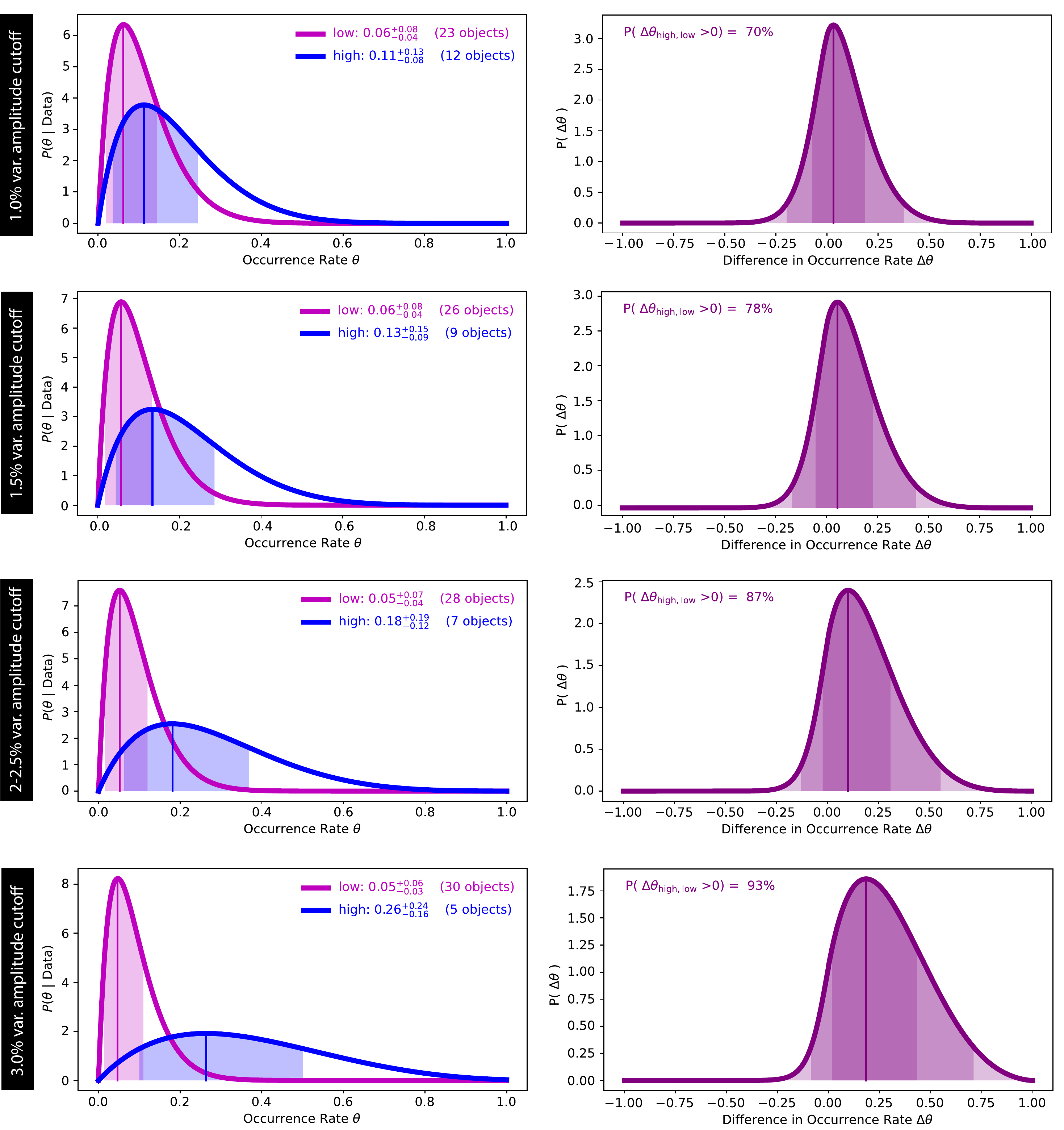}
    \caption{(Left) Quiescent radio occurrence rate distributions for L dwarfs with low versus high photometric variability amplitudes.  Shaded regions show the 68.3\% credible intervals.  Distributions are calculated using the \citet{kao2020} framework for amplitude cutoffs between [1.0\%, 3\%]. The 2\% and 2.5\% cutoff cases are the same, as there were no objects with photometric variability amplitudes between 2 -- 2.5\%. The maximum-likelihood occurrence rate remains approximately constant for the low amplitude samples regardless of the amplitude cutoff, while the high-amplitude occurrence rate appears to increase with increasing amplitude cutoff.  However, this is an artifact of sample size. 
    (Right) Probability density distributions for the difference in occurrence rates $\Delta\theta$ between high and low amplitude samples.  Shaded regions correspond to 68.3\%, 95.5\% and 99.5\% credible intervals. In all cases, we cannot determine if high-amplitude O/IR variability traces radio magnetic activity as our results do not suggest a difference in the radio occurrence rates between high- and low-amplitude variability. } \label{fig:orc}
\end{figure*}

Figure \ref{fig:orc} shows the probability density distributions for the quiescent radio occurrence rate of high-amplitude versus low-amplitude objects for different O/IR amplitude cutoffs.   We also calculate the probability $P(\Delta \theta)$ that the two samples have a difference occurrence rate $\Delta \theta$.  The maximum-likelihood occurrence rate increases with increasing photometric variability amplitude cutoff.  However, an interpretation of this tentative trend requires an abundance of caution, on which we elaborate in \S\ref{orc_discussion}.  Furthermore, Figure \ref{fig:orc} shows that for all variability amplitudes, our results do not suggest a difference in the radio occurrence rates between high- and low-amplitude variability. In all cases, the occurrence rate for the low-amplitude variable objects remains constant at 5 -- 6\%.

\section{Discussion}\label{sec:discussion}

\citet{Kao2016} demonstrated that H$\alpha$ and/or O/IR variability trace radio aurorae, and consequently the quiescent radio emission that accompany all instances of radio aurorae, on L and T dwarfs.  \citet{miles-paez2017} showed that O/IR variability does not trace H$\alpha$ emission. This is unsurprising, since a significant portion of O/IR variability can be attributed to clouds. Therefore, our work asks whether magnetism, traced by radio emission, enhances O/IR variability.

In this work, we isolated the selection effects of H$\alpha$ and O/IR variability by focusing on objects with the latter.
In contrast to the pilot sample from \citet{Kao2016} in which the authors saw a detection rate of 80\%, we see detections in Stokes $I$ and Stokes $V$ in only one of our 17 targets (6\%). Our detection rate is consistent with volume-limited radio surveys that do not bias their target sample with other possible tracers of aurorae  \citep{route2012, Route2013, antonova2013, Lynch2016}.

We must consider the possibility that our observed radio activity detection rate may be a lower limit to the true occurrence rate.
For the 16 objects for which no emission was observed, we consider two possibilities that affect observational completeness, including for both quiescent or flaring emission.  

%\begin{enumerate}
    
First, we may have not observed these targets during a flare. For any target with a rotation period longer than 2 hours, we were not able to observe full coverage of the brown dwarf and thus may have missed when the pulsed emission was beamed towards Earth.  However, since quiescent radio emission at 4 -- 8 GHz \citep{kao2019} accompanies all known examples of ECM emission from ultracool dwarfs at GHz frequencies.  Since we do not detect such emission, these objects likely do not have time-variable ECM emission at our observed frequencies.  Long-term monitoring that provides full phase coverage may prove otherwise. 

Second, the quiescent emission may be too faint to detect.  However, our sensitivities are sufficient to detect quiescent emission for objects emitting at quiescent flux densities that have been observed on L dwarfs, ranging from $[L_{\nu}] \approx 12.6-13.6$ erg s$^{-1}$ Hz$^{-1}$ \citep{kao2020}. Therefore, the possibility of quiescent emission that is too faint to be detectable can most likely be ruled out for our sample.  Furthermore, the occurrence rate calculation takes observational completeness into account.

    % \item These objects may inherently produce no ECM emission and will not be observable at any frequency range.  \citet{kao2019} discuss this possibility in detail.
    
    % \item These objects produce ECM emission either above or below the frequency range observed in this program. Multiple auroral brown dwarfs have been observed in the 8 -- 12 GHz range, and while none have currently been detected below 4 GHz for brown dwarfs, \citet{Villadsen2019} report flaring emission in late M dwarfs at 224 -- 482 MHz that may be attributable to ECM mechanisms. Furthermore, \citet{Vedantham2020} detect possible  100 MHz ECM emission around a late M dwarf that has historically been quiescent in magnetic activity indicators.  Observations at lower radio frequencies will rule in or out this possibility.}  Note that since 2M1750-00 produces emission at the entire 4 -- 8 GHz range of our observations, we expect this object likely emits both above 8 GHz and below 4 GHz.
    
    % \item The auroral activity may be variable, as discussed in greater detail in \citet{kao2019}.  Long-term monitoring will shed insight into this possibility.}
    
%\end{enumerate}

We conclude that O/IR variability by itself does not trace aurorae.

%%%%%%%%%%%%%%%%%%%

\subsection{Occurrence Rates of Quiescent Radio Emission}\label{orc_discussion}
If indeed high-amplitude O/IR variability does not trace quiescent radio emission,  we expect the low- and high-amplitude maximum-likelihood occurrence rates to be similar and to not change with varying amplitude cutoff.  Conversely, if high-amplitude O/IR variability does trace quiescent radio emission, we expect maximum-likelihood occurrence rates between the two samples to diverge with increasing photometric amplitude cutoff up until the true physical amplitude cutoff that identifies the onset of magnetically driven O/IR variability. Using amplitude cutoffs that are not the true cutoff will result in cross-contaminated samples that reduce the distinction between the two samples.

 %This behavior would suggest that localized magnetic heating due to aurorae is likely not a major contributor to high-amplitude variability.

Even though the maximum-likelihood occurrence rate seems to increase with increasing photometric variability amplitude cutoff (Figure \ref{fig:orc}), this behavior is a consequence of decreasing sample sizes.  The detection rate for a given sample determines the lower bound of the maximum-likelihood occurrence rate.  As sample size $N$ decreases, the resolution $1/N$ for detection rates grows.  This pushes the maximum-likelihood occurrence rate higher even if the number of detected objects remains the same in each sample. This is the case for our calculations, in which each of the high-amplitude samples that we define with  various amplitude cutoffs contain the same single radio-emitting object. Thus, we conclude that the tentative rising radio occurrence rate trend that we observe is most likely an artifact of small sample sizes.

Based on our results, we conclude that observed O/IR variability does not trace radio magnetic activity, as the low radio occurrence rates of both the low- and high-amplitude variability samples are consistent both with each other and with the overall L dwarf population from \citet{kao2020}. Comparing this to the prevalence of photometric variability, we infer that optical and infrared variability seen on L dwarfs from 0.5 -- 4.5 $\mu$m is likely predominantly due to cloud phenomena.

However, we also consider other possible explanations for the non-distinct occurrence rates that we observe between our low- and high-amplitude samples:

% degeneracy.  don't know if intrinsic low-ampltidue objects or if they're high amplitude objects that might be inclined in a way.   low amplitude sample contaminated with high amplitude 

 One possibility is geometry.  High inclination objects (equator-on) exhibit higher $J$-band variability amplitudes, with amplitudes strongly attenuated at lower inclinations  \citep{Vos2017}. However, few brown dwarfs have measured inclination angles and existing measurements are not well constrained \citep{Vos2020}.  Consequently,  our low-amplitude sample may be contaminated by high-amplitude objects at low inclinations. This would cause the radio occurrence rate of the low-amplitude sample to shift toward the high-amplitude occurrence rate, since we do not anticipate geometric effects to affect the quiescent radio occurrence rate.   The occurrence rate framework from \citet{kao2020} considers the non-pulsing quiescent radio component rather than the highly beamed auroral component \citep{kao2020}.  Spectral indices measured for brown dwarf quiescent radio emission indicate a gyrosynchrotron mechanism \citep{Williams2015Alma}. While gyrosynchrotron emission from individual electrons is weakly beamed,  observed brown dwarf quiescent emission likely originates from a magnetospheric population of electrons \citep{pineda2017, kao2019, kao2020}. We therefore expect that the velocity distribution of such a population of electrons will smear out the beaming from individual electrons.  Measuring L and T dwarf inclination angles and incorporating the inclination angle dependence into a future study of radio emission on IR/variable L and T dwarfs will rule in or out geometric effects. 

Additionally, we cannot rule out a connection between variability at longer IR wavelengths and radio emission. Quiescent radio emission correlates with markers of auroral activity in ultracool M, L and T dwarfs \citep{pineda2017} that trace strong, kilogauss magnetic fields \citep{Hallinan2008, route2012, Kao2016, kao2018} that may interact with the upper atmospheres of these objects \citep{hallinan2015, pineda2017}.  Magnetic spot heating occurring near the top of the atmosphere may manifest as variability at longer wavelengths, with most flux differences occurring between  2 -- 4 $\mu$m and 5 -- 9 $\mu$m  \citep{morley2014, robinson2014}. While  brown dwarf variability searches typically include the \textit{Spitzer} IRAC channels 1 and 2 at 3.6 $\mu$m and 4.5 $\mu$m, respectively, targeted studies for photometric variability amongst all L dwarfs for which we see pulsed radio emission have searched only from 0.5 -- 2.5 $\mu$m, probing the bottom layers of the bodies' atmospheres at 10 bar and higher \citep{robinson2014}.

Combining detections of radio aurorae implying strong magnetic fields and electron currents with studies at longer amplitudes will allow us to characterize if and how wavelength-dependent variability traces or rules out magnetic spot heating. Multi-wavelength studies of brown dwarfs with pulsed radio emission will be prime targets for \textit{JWST}'s NIRCAM (0.6 -- 5 $\mu$m) and MIRI (5.6 -- 25.5 $\mu$m) instruments.

%\mkao{Our results suggest that enlarging the high- and low-amplitude sample size is essential.  }

% \citet{kao2019} suggest that the discovery of pulsed} radio emission from the high-amplitude variable brown dwarf SIMP J01365662+0933473 points to the possibility of a variability-inducing mechanism in addition to patchy clouds occurring in high-amplitude variable objects.  This interpretation is consistent with the  tentatively higher quiescent} radio occurrence rate in our high amplitude sample.

%\subsection{Photometric Variability at O/IR Wavelengths Not Due to Auroral Spot Heating}

\subsection{Auroral tracers: H$\alpha$ emission or IR variability? }

\begin{figure}[t]
    \centering
    \includegraphics[width=\linewidth]{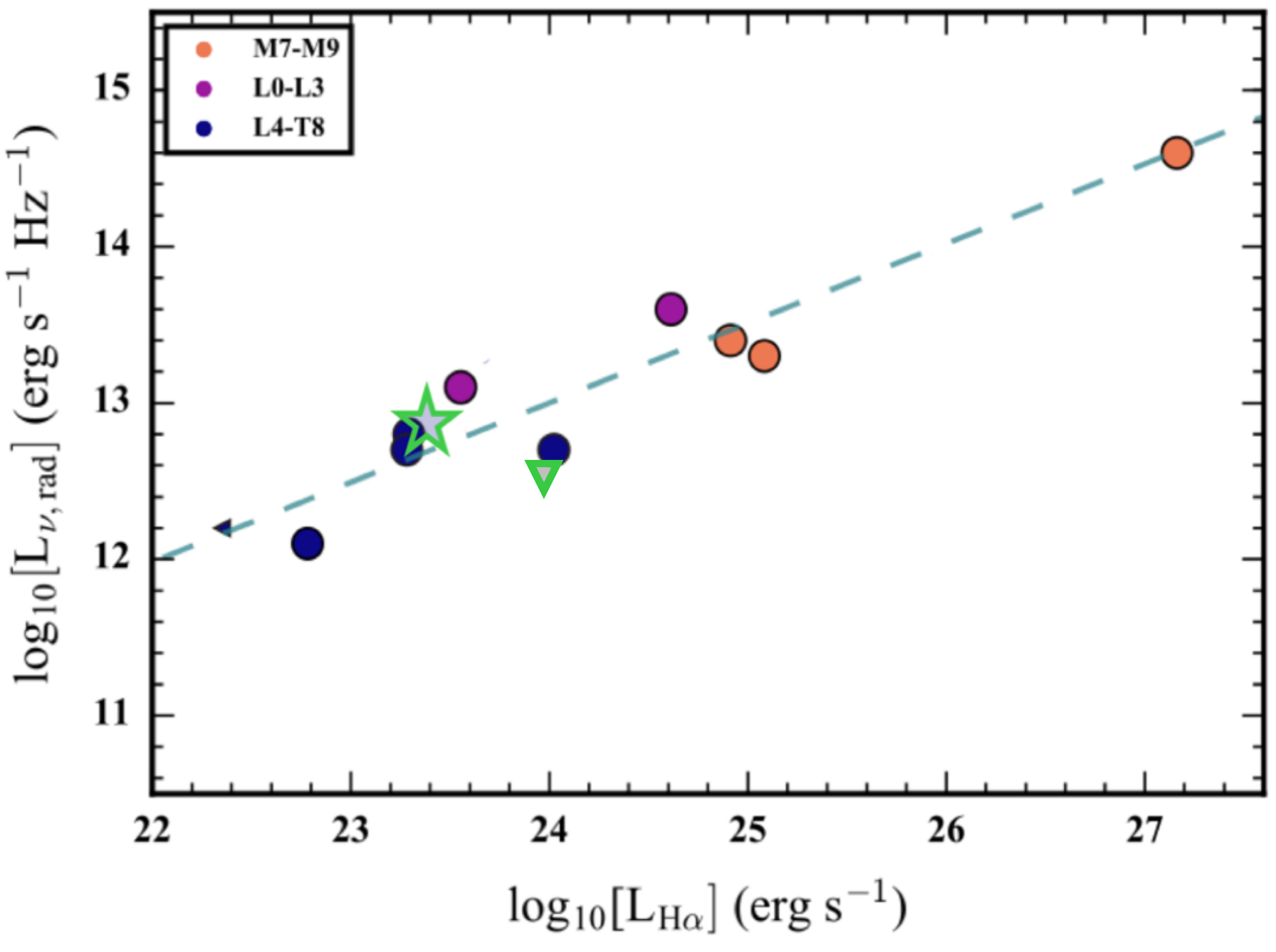}
    \caption{H$\alpha$ luminosity compared to radio luminosity for pulsed radio emitters from \citet{pineda2017}. The addition of 2M1750-00 is represented by a green star and follows the known relationship well. DENIS 1058-15, the only other target with an H$\alpha$ measurement, is also shown as a triangle representing the radio upper limit.}
    \label{fig:halpha}
\end{figure}

For 2M1750-00, we observe a coherent ECM flare that is characterized by a broad peak in emission with additional substructure. 
Interestingly, 2M1750-00 additionally has measured H$\alpha$ emission of $f_{\alpha} = 21.4 \pm 4.8$ x 10$^{-18}$ erg s$^{-1}$ cm$^{-2}$ \citep{Pineda2016}. This was one of two objects in our sample for which H$\alpha$ emission has been detected, with the caveat that several of the targets have not yet been observed for H$\alpha$ emission. \citet{pineda2017} demonstrated a tight correlation between H$\alpha$ and quiescent radio luminosities among pulsed radio emitters \citep{pineda2017}, and we show in Figure \ref{fig:halpha} that our detected target directly follows this relationship. If this relationship holds, then the other target in our sample with measured H$\alpha$ emission, DENIS 1058-15, may be 
%may not have been detected for reasons 2 -- 4 discussed in the opening paragraphs of this section and would thus make 
a good target for future follow-up observations.

Combining our results with those of \citet{Kao2016}, we suggest that  H$\alpha$ in the spectra of a brown dwarfs regardless of its temperature points to non-thermal magnetic processes; furthermore, in late L dwarfs and T-dwarfs it is a reliable sign of auroral currents. This is unsurprising, since H$\alpha$ has long been seen as an indicator of magnetic activity in the chromosphere of stars \citep[e.g.][]{Linsky1982, Walkowicz2008}. Moreover, in the cooler atmospheres of brown dwarfs, H$\alpha$ emission has been seen to decline rapidly \citep{berger2010, williams2014, schmidt2015}, signaling the separation between stellar chromospheric magnetic activity and substellar magnetospheric activity. \citet{Pineda2016} found that the detection rate of H$\alpha$ in brown dwarfs L4 and later was 9.2\%, which is consistent with the the putative quiescent radio occurrence rate for L  dwarfs \citep{kao2020}.  \citet{Pineda2016} proposed that a possible connection between H$\alpha$ emission and auroral activity could be through the raining down of electrons via flux tubes between a brown dwarf and an inferred satellite. Such a situation would mimic that of Io's auroral footprint on Jupiter \citep[e.g.][]{Vasavada1999}. Therefore, it remains a possibility that there are yet undetected companions to brown dwarfs that exhibit H$\alpha$ emission.

\subsection{T Dwarf Aurorae and Photometric Variability}

The results of this work will soon be combined with a similar study of the relationship between T Dwarf aurorae and O/IR variability to yield a complete picture throughout the range of brown dwarf spectral types through which radio aurorae have been observed (Kao et al., in prep). 

Because T dwarfs have different atmospheric compositions due to their cooler temperatures, the structure of the thermal profile and the ability for atmospheric circulation and transport may differ. \citet{morley2014} showed that flux ratios from excess emission due to spot heating at various atmospheric depths both increases and shifts redward as objects cool from 1000 K to 400 K. This suggests that magnetic spot heating may cause stronger photometric responses in T dwarfs.

\section{Conclusions} \label{sec:conclusions}

We searched over 2 hours of observations for quiescent and/or pulsed radio emission in 17 L dwarfs from 4 -- 8 GHz. We observe highly circularly polarized, pulsed emission in only one target, 2M1750-00. Additionally, 2M1750-00 was the only object for which we observed quiescent radio emission, furthering the evidence that quiescent emission and auroral emission are related. We determine a lower limit on the magnetic field strength of 2M1750-00 of 2.9 kG. 

We selected our sample for clear O/IR variability. Because we did not see a detection rate much greater than that of previous volume-limited samples, we infer that auroral magnetic activity does not play a role in the O/IR variability observed on these targets. The depth at which auroral magnetic activity may influence the atmosphere is not constrained, so observations at longer wavelengths that probe deeper into brown dwarf atmospheres may indeed show such a connection.

Our empirical results are supported by a theoretical framework to calculate the occurrence rate distributions of quiescent radio activity for brown dwarfs with low- and high-amplitude variability, based on the maximum-likelihood occurrence rate framework from \citet{kao2020}. We find that the occurrence rates of quiescent emission in L dwarfs with low- and high-amplitude variability are between 5 -- 6\% and 11 -- 26\%, respectively, depending on the assumed cutoff between low- and high-amplitude variability.  As we increased the amplitude cutoff from 1\% to 3\%, occurrence rate of the low-amplitude sample remained relatively constant, while the occurrence rate increased with increasing amplitude cutoff for the high-amplitude sample. However, we determine that this is an artifact of sample sizes and conclude that high amplitude O/IR variability does not trace radio magnetic activity in L dwarfs. 
Future studies improving and expanding upon inclination measurements of brown dwarfs together with studies of IR variability beyond 5$\mu$m will aid in forming a more thorough assessment of a relationship between brown dwarf photometric variability and radio magnetic activity. 

%A larger statistical radio survey is necessary to determine the possibility that the energy deposition from the auroral currents contributes to high amplitude atmospheric  variability. 

Finally, we find that the only radio-bright object in our sample, 2M1750-00, is also an H$\alpha$ emitter.   We show that its quiescent radio luminosity is consistent with an existing correlation between H$\alpha$ luminosities and quiescent radio luminosities in auroral ultracool dwarfs.  We conclude that 
%This object is one of two objects in our sample ta
%2M1750-00 was one of two objects in our sample that showed H$\alpha$ emission in addition to variability.
%While our sample was selected for only O/IR variability with a detection rate of 6\%, the sample of \citet{Kao2016} was selected for O/IR variability as well as H$\alpha$ emission and saw a detection rate of 80\%. We conclude that 
H$\alpha$ emission in the spectra of brown dwarfs is the stronger indicator of strong magnetic fields traced by radio emission.
%active radio auroral regions. 

%\vspace{0.5cm}

\acknowledgments
The authors would like to thank the anonymous referee for an insightful report. T.R.Y. and M.M.K. would like to thank Cameron Voloshin for consulting on the statistics of this study. T.R.Y. would additionally like to thank Danny Jacobs, Adam Beardsley, and Judd Bowman for useful discussions and CASA help. Support for this work was provided by NASA through the NASA Hubble Fellowship grant HST-HF2-51411.001-A awarded by the Space Telescope Science Institute, which is operated by the Association of Universities for Research in Astronomy, Inc., for NASA, under contract NAS5-26555; and by the National Radio Astronomy Observatory. The National Radio Astronomy Observatory is a facility of the National Science Foundation operated under cooperative agreement by Associated Universities, Inc. This work is based on observations made with the NSF's Karl G. Jansky Very Large Array (VLA). This research has made use of the SIMBAD and VizieR databases, operated at CDS, Strasbourg, France; and the European Space Agency (ESA) mission \textit{Gaia} (\url{https://www. cosmos.esa.int/gaia}), processed by the Gaia Data Processing and Analysis Consortium (DPAC, \url{https://www. cosmos.esa.int/web/gaia/dpac/consortium}). 

\software{CASA \, \citep{CASA}, Astropy \, \citep{astropy:2018},\, Matplotlib \citep{matplotlib},\, Numpy \,\citep{numpy2}, \, Scipy \, \citep{scipy}, \, MATLAB \, \citep{MATLAB:2018}.}

\bibliography{bibliograph}

\begin{longrotatetable}
\begin{deluxetable}{l c c c c l l}
\centering
\tabletypesize{\small}
\tablecaption{\normalsize{Variability information of the L dwarfs in our sample. Amplitude units are in magnitudes unless otherwise noted.}\label{tab:target_var_info}} 
\tablehead{
\colhead{Name} \vspace{-0.2cm} & \colhead{Filter} & \colhead{Amp.} & \colhead{\% P2P} & \colhead{Periodic} & \colhead{Ref.} & \colhead{Note}\\
\vspace{-0.2cm}
}
\startdata
2MASSI J0030300-145033	&	Ks	&	0.19 $\pm$ 0.11	&	1.32	&	yes	&	EBB03	&	Ks = 14.38 $\pm$ 0.08	\\
	&	JHKs	&	0	&	0	&	---	&	KO05	&		\\
	&	JHKs	&	0	&	0	&	---	&	CL08	&		\\
	&	izJ	&	0	&	0	&	---	&	SC15	&		\\
	&	J	&	0	&	0	&	---	&	RA14b	&		\\
2MASSI J0103320+193536	&	3.6 $\mu$m	&	0.56 $\pm$ 0.03 \%	&	0.56 $\pm$ 0.03	&	yes	&	ME15	&		\\
	&	4.5 $\mu$m	&	0.87 $\pm$ 0.09 \%	&	0.87 $\pm$ 0.09	&	yes	&	ME15	&		\\
	&	Ks	&	0.10 $\pm$0.02	&	0.71	&	yes	&	EBB03	&	KS = 14.15 $\pm$ 0.07	\\
	&	Js	&	0	&	0	&	---	&	VO18	&		\\
2MASS J01075242+0041563	&	3.6 $\mu$m	&	1.27 $\pm$ 0.13 \%	&	1.27 $\pm$ 0.13 	&	no	&	ME15	&		\\
	&	4.5 $\mu$m	&	1.0 $\pm$ 0.2 \%	&	1.0 $\pm$ 0.2 	&	no	&	ME15	&		\\
2MASSW J0310599+164816	&	J	&	2 \%/hr	&	1.5	&	unknown	&	BU14	&	Assuming P$_{\text{rot}}$ = 2x observation length	\\
2MASS J08354256-0819237	&	I	&	0.01	&	0.06	&	yes	&	KO04a	&	I = 17.6	\\
	&	Js	&	1.3 $\pm$ 0.2 \%	&	1.3 $\pm$ 0.2	&	yes	&	RA14b	&		\\
	&	Rc	&	0	&	0	&	---	&	KO05	&		\\
	&	Js	&	1.6 $\pm$ 0.5 \%	&	1.6 $\pm$ 0.5	&	yes	&	WI14	&		\\
	&	0.9 $\mu$m	&	0	&	0	&	---	&	SC17	&		\\
	&	0.96 $\mu$m	&	0	&	0	&	---	&	SC17	&		\\
	&	1.02 $\mu$m	&	0	&	0	&	---	&	SC17	&		\\
	&	1.08 $\mu$m	&	0	&	0	&	---	&	SC17	&		\\
	&	1.14 $\mu$m	&	0	&	0	&	---	&	SC17	&		\\
	&	1.20$\mu$m	&	0	&	0	&	---	&	SC17	&		\\
	&	1.26 $\mu$m	&	0	&	0	&	---	&	SC17	&		\\
	&	1.33 $\mu$m	&	0	&	0	&	---	&	SC17	&		\\
	&	1.39 $\mu$m	&	0	&	0	&	---	&	SC17	&		\\
	&	1.45 $\mu$m	&	0	&	0	&	---	&	SC17	&		\\
	&	1.51 $\mu$m	&	0	&	0	&	---	&	SC17	&		\\
	&	1.57 $\mu$m	&	0	&	0	&	---	&	SC17	&		\\
	&	1.63 $\mu$m	&	0	&	0	&	---	&	SC17	&		\\
	&	1.69 $\mu$m	&	0	&	0	&	---	&	SC17	&		\\
	&	1.76 $\mu$m	&	0	&	0	&	---	&	SC17	&		\\
	&	1.82 $\mu$m	&	0	&	0	&	---	&	SC17	&		\\
	&	1.88 $\mu$m	&	0	&	0	&	---	&	SC17	&		\\
	&	1.94 $\mu$m	&	0	&	0	&	---	&	SC17	&		\\
	&	2.00 $\mu$m	&	0	&	0	&	---	&	SC17	&		\\
	&	2.06 $\mu$m	&	0	&	0	&	---	&	SC17	&		\\
	&	2.12 $\mu$m	&	0	&	0	&	---	&	SC17	&		\\
	&	2.19 $\mu$m	&	0	&	0	&	---	&	SC17	&		\\
	&	2.25 $\mu$m	&	0	&	0	&	---	&	SC17	&		\\
	&	2.31 $\mu$m	&	0	&	0	&	---	&	SC17	&		\\
	&	2.37 $\mu$m	&	0	&	0	&	---	&	SC17	&		\\
2MASS J10101480-0406499	&	Js	&	3.6 $\pm$ 0.4 \%	&	3.6 $\pm$ 0.4	&	yes	&	RA14b	&		\\
	&	Js	&	5.1 $\pm$ 1.1 \%	&	5.1 $\pm$ 1.1	&	yes	&	WI14	&		\\
2MASS J10433508+1213149	&	3.6 $\mu$m	&	1.54 $\pm$ 0.15 \%	&	1.54 $\pm$ 0.15	&	no	&	ME15	&		\\
	&	4.5 $\mu$m	&	1.2 $\pm$ 0.2 \%	&	1.2 $\pm$ 0.2	&	no	&	ME15	&		\\
2MASS J10584787-1548172	&	3.6 $\mu$m	&	0.39 $\pm$ 0.04 \%	&	0.39 $\pm$ 0.04	&	yes	&	ME15	&		\\
	&	4.5 $\mu$m	&	0	&	0	&	---	&	ME15	&		\\
	&	I	&	0	&	0	&	---	&	KO13	&		\\
	&	R	&	0	&	0	&	---	&	KO13	&		\\
	&	3.6 $\mu$m	&	0.388 $\pm$ 0.043 \%	&	0.388 $\pm$ 0.043	&	yes	&	HE13	&		\\
	&	4.5 $\mu$m	&	0.090 $\pm$ 0.056 \%	&	0.090 $\pm$ 0.056	&	yes	&	HE13	&		\\
	&	J	&	0.843 $\pm$ 0.098 \%	&	0.843 $\pm$ 0.098	&	yes	&	HE13	&		\\
2MASS J12195156+3128497	&	J	&	$\sim$ 3 -- 6 \%/hr	&	$\sim$ 2.5 -- 5	&	unknown	&	BU14	&		\\
	&	H	&	0	&	0	&	---	&	BU14	&		\\
2MASS J14252798-3650229	&	J	&	0.6 $\pm$ 0.1 \%	&	0.6 $\pm$ 0.1	&	yes	&	RA14b	&		\\
	&	Js	&	0.7 $\pm$ 0.3  \%	&	0.7 $\pm$ 0.3 	&	yes	&	VO18	&		\\
2MASS J16154255+4953211	&	3.6 $\mu$m	&	0.9 $\pm$ 0.2 \%	&	0.9 $\pm$ 0.2	&	yes	&	ME15	&		\\
	&	4.5 $\mu$m	&	0	&	0	&	---	&	ME15	&		\\
	&	J	&	0	&	0	&	---	&	VO18	&		\\
2MASS J16322911+1904407	&	J	&	0	&	0	&	---	&	BU14	&		\\
	&	H	&	0	&	0	&	---	&	BU14	&		\\
	&	Js	&	0	&	0	&	---	&	WI14	&		\\
	&	3.6 $\mu$m	&	0.42 $\pm$ 0.08 \%	&	0.42 $\pm$ 0.08	&	yes	&	ME15	&		\\
	&	4.5 $\mu$m	&	0.5 $\pm$ 0.3 \%	&	0.5 $\pm$ 0.3	&	yes	&	ME15	&		\\
2MASS J17114573+2232044	&	J	&	0.206 $\pm$ 0.041	&	1.21	&	yes	&	KH13	&	J = 17.09 $\pm$ 0.18	\\
	&	K'	&	1.186 $\pm$ 0.083	&	8.05	&	yes	&	KH13	&	K = 14.73 $\pm$ 0.10; Comparison stars not stable	\\
	&	J	&	0	&	0	&	---	&	BU14	&		\\
2MASSI J1721039+334415	&	3.6 $\mu$m	&	0.33 $\pm$ 0.07 \%	&	0.33 $\pm$ 0.07	&	yes	&	ME15	&		\\
	&	4.5 $\mu$m	&	0	&	0	&	---	&	ME15	&		\\
2MASS J17502484-0016151	&	I	&	0	&	0	&	---	&	KO13	&		\\
	&	R	&	0	&	0	&	---	&	KO13	&		\\
	&	J	&	0.75 \%/hr	&	0.5	&	unknown	&	BU14	&	Assuming P$_{\text{rot}}$ = 2x observation length	\\
	&	H	&	0	&	0	&	---	&	BU14	&		\\
	&	J	&	0	&	0	&	---	&	RA14b	&		\\
2MASS J18212815+1414010	&	I	&	0	&	0	&	---	&	KO13	&		\\
	&	R	&	0	&	0	&	---	&	KO13	&		\\
	&	3.6 $\mu$m	&	0.54 $\pm$ 0.05 \%	&	0.54 $\pm$ 0.05	&	yes	&	ME15	&		\\
	&	4.5 $\mu$m	&	0.71 $\pm$ 0.14 \%	&	0.71 $\pm$ 0.14	&	yes	&	ME15	&		\\
	&	1.1 -- 1.7 $\mu$m	&	1.77 $\pm$ 0.11 \%	&	1.77 $\pm$ 0.11	&	unknown	&	YA15	&		\\
	&	1.4 $\mu$m	&	1.54 $\pm$ 0.21 \%	&	1.54 $\pm$ 0.21	&	unknown	&	YA15	&		\\
	&	0.9 $\mu$m	&	0	&	3	&	yes	&	SC17	&		\\
	&	0.96 $\mu$m	&	0	&	2.5	&	yes	&	SC17	&		\\
	&	1.02 $\mu$m	&	0	&	2.5	&	yes	&	SC17	&		\\
	&	1.08 $\mu$m	&	0	&	1.8	&	yes	&	SC17	&		\\
	&	1.14 $\mu$m	&	0	&	1.8	&	yes	&	SC17	&		\\
	&	1.20$\mu$m	&	0	&	1.5	&	yes	&	SC17	&		\\
	&	1.26 $\mu$m	&	0	&	1.3	&	yes	&	SC17	&		\\
	&	1.33 $\mu$m	&	0	&	1.3	&	yes	&	SC17	&		\\
	&	1.39 $\mu$m	&	0	&	1.3	&	yes	&	SC17	&		\\
	&	1.45 $\mu$m	&	0	&	1.3	&	yes	&	SC17	&		\\
	&	1.51 $\mu$m	&	0	&	0.9	&	yes	&	SC17	&		\\
	&	1.57 $\mu$m	&	0	&	0.8	&	yes	&	SC17	&		\\
	&	1.63 $\mu$m	&	0	&	0.7	&	yes	&	SC17	&		\\
	&	1.69 $\mu$m	&	0	&	0.5	&	yes	&	SC17	&		\\
	&	1.76 $\mu$m	&	0	&	0	&	---	&	SC17	&		\\
	&	1.82 $\mu$m	&	0	&	0	&	---	&	SC17	&		\\
	&	1.88 $\mu$m	&	0	&	0	&	---	&	SC17	&		\\
	&	1.94 $\mu$m	&	0	&	0.6	&	yes	&	SC17	&		\\
	&	2.00 $\mu$m	&	0	&	0	&	---	&	SC17	&		\\
	&	2.06 $\mu$m	&	0	&	0	&	---	&	SC17	&		\\
	&	2.12 $\mu$m	&	0	&	0	&	---	&	SC17	&		\\
	&	2.19 $\mu$m	&	0	&	0	&	---	&	SC17	&		\\
	&	2.25 $\mu$m	&	0	&	0	&	---	&	SC17	&		\\
	&	2.31 $\mu$m	&	0	&	0	&	---	&	SC17	&		\\
	&	2.37 $\mu$m	&	0	&	0	&	---	&	SC17	&		\\
2MASS J21481628+4003593	&	J	&	0	&	0	&	---	&	KH13	&		\\
	&	3.6 $\mu$m	&	1.33 $\pm$ 0.07 \%	&	1.33 $\pm$ 0.07	&	yes	&	ME15	&		\\
	&	4.5 $\mu$m	&	1.03 $\pm$ 0.1 \%	&	1.03 $\pm$ 0.1	&	yes	&	ME15	&		\\
\enddata
\tablerefs{
(BU14)		\citet{buenzli2014};
(CL08)		\citet{clarke2008};
(EBB03)		\citet{enoch2003};
(HE13)		\citet{Heinze2013};
(KH13)		\citet{Khandrika2013};
(KO04a)		\citet{koen2004a};
(KO13)		\citet{Koen2013};
(KO05)		\citet{Koen2005};
(ME15)		\citet{Metchev2015};
(RA14b)		\citet{Radigan2014};
(SC17)		\citet{Schlawin2017};
(SC15)		\citet{schmidt2015};
(VO18)		\citet{Vos2018};
(WI14)		\citet{wilson2014};
(YA15)		\citet{yang2015};
}
\end{deluxetable}
\end{longrotatetable}
\begin{longrotatetable}
\begin{deluxetable}{l c c c h c c c c c c l l}
\tabletypesize{\tiny}
\tablecaption{\normalsize{Variability information of L dwarfs that have been targeted by radio searches to search for quiescent emission.} We include only spectral types $>$L2.5 to match those of our target sample. Amplitude units are in magnitudes unless otherwise noted. Radio fluxes are reported in $\mu$Jy and frequencies in GHz.}\label{tab:var_info}
\tablehead{
 \colhead{Name} \vspace{-0.2cm} & \colhead{SpT} & \colhead{Ref.} & \colhead{$f_{I}$\tablenotemark{a}} & \colhead{} & \colhead{Ref.} & \colhead{Var.} & \colhead{Filter} & \colhead{Amp.} & \colhead{\% P2P} & \colhead{Periodic} & \colhead{Ref.} & \colhead{Note}\\
}
\startdata
2MASS J00043484-4044058	&	L5+L5	&	RE08	&	100	&	5.5	&	LY16	&	---	&	---	&	---	&	---	&	---	&	---	&	No info	\\
2MASS J00303013-1450333 	&	L7V	&	KI00	&	<17.4	&	4 -- 8	&	This work	&	yes	&	Ks	&	0.2	&	1.39	&	yes	&	EBB03	&		\\
	&		&		&		&		&		&	yes	&	Ks	&	0.19 $\pm$ 0.11	&	1.32	&	yes	&	EBB03	&	Ks = 14.38 $\pm$ 0.08	\\
	&		&		&		&		&		&	no	&	JHK	&	0	&	0	&	---	&	KT05	&		\\
	&		&		&		&		&		&	no	&	J	&	0	&	0	&	---	&	RA14a	&		\\
	&		&		&		&		&		&	no	&	J	&	0	&	0	&	---	&	SC15	&		\\
	&		&		&		&		&		&	no	&	J	&	0	&	0	&	---	&	CL08	&		\\
2MASS J00325937+1410371 	&	L9	&	SC14	&	<1101	&	4.75	&	RW13	&	---	&	---	&	---	&	---	&	---	&	---	&	No info	\\
2MASS J00361617+1821104	&	L3.5	&	RE00	&	152	&	4.86	&	BE05	&	no	&	IR	&	0	&	0	&	---	&	GU09	&		\\
	&		&		&		&		&		&	yes	&	I	&	0.015	&	0.09	&	yes	&	LA07	&		\\
	&		&		&		&		&		&	yes	&	I	&	0.01	&	0.05	&	no	&	MA07	&	I/R variaibllity anti-correlated	\\
	&		&		&		&		&		&	yes	&	R	&	0.03	&	0.19	&	no	&	MA07	&	I/R variaibllity anti-correlated	\\
	&		&		&		&		&		&	yes	&	I	&	0.016	&	0.1	&	no	&	KO13	&	long-term var. possible, >56 mmag amp.	\\
	&		&		&		&		&		&	no	&	R	&	0.007	&	0.04	&	no	&	KO13	&		\\
	&		&		&		&		&		&	no	&	I	&	0	&	0	&	---	&	GE02 	&		\\
	&		&		&		&		&		&	yes	&	R	&	3.40 $\pm$ 0.11 \%	&	3.4	&	yes	&	CR16	&		\\
	&		&		&		&		&		&	yes	&	I	&	2.11 $\pm$ 0.09 \%	&	2.11	&	yes	&	CR16	&		\\
	&		&		&		&		&		&	yes	&	z	&	2.74 $\pm$ 0.08 \%	&	2.74	&	yes	&	CR16	&		\\
	&		&		&		&		&		&	yes	&	J	&	1.22 $\pm$ 0.04 \%	&	1.22	&	yes	&	CR16	&		\\
	&		&		&		&		&		&	yes	&	H	&	0.45 $\pm$ 0.05 \%	&	0.45	&	yes	&	CR16	&		\\
	&		&		&		&		&		&	yes	&	Ks	&	1.07 $\pm$ 0.08 \%	&	1.07	&	yes	&	CR16	&		\\
	&		&		&		&		&		&	yes	&	all - RIzJHKs	&	1.36 $\pm$ 0.03 \%	&	1.36	&	yes	&	CR16	&		\\
	&		&		&		&		&		&	yes	&	3.6$\mu$m	&	0.47 $\pm$ 0.05 \%	&	0.47	&	no	&	ME15	&		\\
	&		&		&		&		&		&	yes	&	4.5$\mu$m	&	0.19 $\pm$ 0.04 \%	&	0.19	&	no	&	ME15	&		\\
	&		&		&		&		&		&	yes	&	I 	&	1.98 -- 2.20 \% 	&	2.2	&	yes	&	HA13 	&		\\
2MASSI J0103320+193536	&	L6V	&	KI00	&	<11.4	&	4 -- 8	&	This work	&	yes	&	3.6$\mu$m	&	0.56 $\pm$ 0.03 \%	&	0.56	&	yes	&	ME15	&		\\
	&		&		&		&		&		&	yes	&	4.5$\mu$m	&	0.87 $\pm$ 0.09 \%	&	0.87	&	yes	&	ME15	&		\\
	&		&		&		&		&		&	yes	&	Ks	&	0.10 $\pm$ 0.02	&	0.71	&	no	&	EBB03	&	Ks = 14.15 $\pm$ 0.07	\\
	&		&		&		&		&		&	no	&	Js	&	0	&	0	&	---	&	VO18	&		\\
2MASS J01075242+0041563	&	L8	&	SC14	&	<10.2	&	4 -- 8	&	This work	&	yes	&	3.6$\mu$m	&	1.27 $\pm$ 0.13 \%	&	1.27	&	no	&	ME15	&		\\
	&		&		&		&		&		&	yes	&	4.5$\mu$m	&	1.0 $\pm$ 0.2 \%	&	1	&	no	&	ME15	&		\\
2MASS J01443536-0716142	&	L6.5	&	SC14	&	<33	&	8.46	&	BE06	&	yes	&	I	&	0.06 \%	&	0.06	&	no	&	KO13	&		\\
2MASS J02050344+1251422	&	L5V	&	KI00	&	<48	&	8.46	&	BE06	&	---	&	---	&	---	&	---	&	---	&	---	&	No info	\\
2MASS J02052940-1159296	&	L7+L7	&	RE06	&	<30	&	8.46	&	BE06	&	---	&	---	&	---	&	---	&	---	&	---	&	No info	\\
2MASS J02511490-0352459	&	L3	&	SC07	&	<36	&	8.46	&	BE06	&	no	&	I	&	0	&	0	&	---	&	KO13	&		\\
	&		&		&		&		&		&	no	&	R	&	0	&	0	&	---	&	KO13	&		\\
2MASS J02550357-4700509	&	L9	&	SC14	&	<30.9	&	5.5	&	LY16	&	no	&	J	&	0	&	0	&	---	&	KT05	&		\\
	&		&		&		&		&		&	no	&	H	&	0	&	0	&	---	&	KT05	&		\\
	&		&		&		&		&		&	no	&	Ks	&	0	&	0	&	---	&	KT05	&		\\
2MASS J02572581-3105523	&	L8.5	&	SC14	&	<63.0	&	9	&	LY16	&	no	&	Ic	&	0	&	0	&	---	&	KO13	&		\\
	&		&		&		&		&		&	no	&	Js	&	0	&	0	&	---	&	WI14	&		\\
	&		&		&		&		&		&	no	&	J	&	0	&	0	&	---	&	RA14a	&		\\
2MASSW J0310599+164816	&	L8V	&	KI00	&	<10.8	&	4 -- 8	&	This work	&	yes	&	J	&	2 \%/hr	&	1.5	&	unknown	&	BU14	&	Assuming P$_{\text{rot}}$ = 2x observation length	\\
2MASS J03261367+2950152	&	L4.6V	&	BG14	&	<1293	&	4.75	&	RW13	&	---	&	---	&	---	&	---	&	---	&	---	&	No info	\\
2MASS J03284265+2302051	&	L9.5	&	SC14	&	<1044	&	4.75	&	RW13	&	yes	&	Ks	&	0.43 $\pm$ 0.16 	&	2.89	&	no	&	EBB03	&	K = 14.87	\\
	&		&		&		&		&		&	no	&	J	&	0	&	0	&	---	&	RA14a	&		\\
	&		&		&		&		&		&	no	&	3.6$\mu$m	&	0	&	0	&	---	&	ME15	&		\\
	&		&		&		&		&		&	no	&	4.5$\mu$m	&	0	&	0	&	---	&	ME15	&		\\
2MASS J03400942-6724051	&	L7	&	FA09	&	<27.0	&	5.5	&	LY16	&	no	&	Js	&	---	&	---	&	---	&	WI14	&		\\
2MASS J03552337+1133437	&	L3-L6	&	GA15b	&	<45	&	4.9	&	AN13	&	no	&	I	&	---	&	---	&	---	&	KO13	&		\\
2MASS J04234858-0414035	&	L6.5+T2	&	DU12	&	54.1	&	4 -- 8	&	KAO16	&	yes	&	Ks	&	0.30 $\pm$ 0.18 	&	2.32	&	no	&	EBB03	&		\\
	&		&		&		&		&		&	no	&	J	&	0.015	&	0	&	---	&	KT05	&		\\
	&		&		&		&		&		&	no	&	H	&	0.011	&	0	&	---	&	KT05	&		\\
	&		&		&		&		&		&	no	&	K	&	0.002	&	0	&	---	&	KT05	&		\\
	&		&		&		&		&		&	yes	&	J	&	0.008 $\pm$ 0.0008	&	0.06	&	yes	&	CL08	&		\\
	&		&		&		&		&		&	no	&	I	&	0	&	0	&	---	&	KO13	&		\\
	&		&		&		&		&		&	no	&	Js	&	0	&	0	&	---	&	WI14	&		\\
2MASS J04390101-2353083	&	L4.5	&	SC14	&	<42	&	8.46	&	BE06	&	yes	&	Js	&	2.6 $\pm$ 0.5 \%	&	2.6	&	yes	&	WI14	&	J = 4.408 $\pm$ 0.029 	\\
	&		&		&		&		&		&	no	&	Ic	&	0	&	0	&	---	&	KO13	&		\\
	&		&		&		&		&		&	no	&	Ic	&	0	&	0	&	---	&	KO05	&		\\
	&		&		&		&		&		&	no	&	Js	&	0	&	0	&	---	&	RA14b	&		\\
2MASS J04455387-3048204	&	L2	&	SC07	&	<66	&	8.46	&	BE06	&	no	&	Ic	&	0	&	0	&	---	&	WI14	&		\\
	&		&		&		&		&		&	no	&	Ic	&	0	&	0	&	---	&	KO13	&		\\
	&		&		&		&		&		&	no	&	Js	&	0	&	0	&	---	&	KO04b	&		\\
2MASS J05002100+0330501	&	L4pec	&	GA15b	&	<51	&	4.9	&	AN13	&	---	&	---	&	---	&	---	&	---	&	---	&	No info	\\
2MASS J05233822-1403022	&	L2.5	&	CR03	&	<39	&	8.46	&	BE06	&	no	&	Ic	&	0	&	0	&	---	&	WI14	&		\\
	&		&		&		&		&		&	no	&	Ic	&	0	&	0	&	---	&	KO13	&		\\
	&		&		&		&		&		&	no	&	Js	&	0	&	0	&	---	&	KT05	&		\\
2MASS J05395200-0059019	&	L5	&	SC14	&	<48	&	4.9	&	AN13	&	no	&	Ic	&	0	&	0	&	---	&	KO13	&		\\
	&		&		&		&		&		&	yes	&	I	&	0.009 \%	&	0.009	&	no	&	BM01	&		\\
	&		&		&		&		&		&	no	&	J	&	0	&	0	&	---	&	BU14	&		\\
	&		&		&		&		&		&	no	&	H	&	0	&	0	&	---	&	BU14	&		\\
WISEP J060738.65+242953.4	&	L9	&	CA13	&	15.6	&	6.05	&	GI16	&	no	&	832 nm	&	0	&	0	&	---	&	GI16 	&		\\
	&		&		&		&		&		&	no	&	3.6$\mu$m	&	0	&	0	&	---	&	GI16 	&		\\
	&		&		&		&		&		&	no	&	4.5$\mu$m	&	0	&	0	&	---	&	GI16 	&		\\
2MASS J06523073+4710348	&	L4.5	&	BUR10	&	<33	&	8.46	&	BE06		---	&	---	&	---	&	---	&	---	&	---	&	No info	\\
2MASS J07003664+3157266	&	L3+L6.5	&	DU12	&	<42	&	4.9	&	AN13	&	---	&	---	&	---	&	---	&	---	&	---	&	No info	\\
2MASS J07464256+2000321	&	L0+L1.5	&	DU17	&	<48	&	8.46	&	BE06	&	yes	&	Cousins I	&	0.07 \%	&	0.07	&	yes	&	GE02	&	I = 15.11	\\
	&		&		&		&		&		&	yes	&	I-band	&	0.46 \%	&	0.46	&	yes	&	CL02	&		\\
	&		&		&		&		&		&	no	&	R band	&	0	&	0	&	---	&	MA07	&		\\
	&		&		&		&		&		&	no	&	I band	&	0	&	0	&	---	&	MA07	&		\\
	&		&		&		&		&		&	yes	&	J	&	0.05 \%	&	0.05	&	unknown	&	BL08	&	J = 11.74	\\
	&		&		&		&		&		&	yes	&	H	&	0.05 \%	&	0.05	&	unknown	&	BL08	&	H = 11.00	\\
	&		&		&		&		&		&	yes	&	Ks	&	0.06 \%	&	0.06	&	unknown	&	BL08	&	Ks = 10.49	\\
	&		&		&		&		&		&	yes	&	Cousins I	&	0.54 \%	&	0.54	&	yes	&	KO13	&	I = 15.11	\\
	&		&		&		&		&		&	no	&	Cousins R	&	0	&	0	&	---	&	KO13	&		\\
	&		&		&		&		&		&	yes	&	I band	&	1.52 \%	&	1.52	&	yes	&	HA13 	&		\\
2MASS J08251968+2115521	&	L7.5V	&	KI00	&	<45	&	8.46	&	BE06	&	yes	&	J	&	1 \%/hr	&	1.51	&	unknown	&	BU14	&	J = 15.10; H = 13.79	\\
	&		&		&		&		&		&	yes	&	H	&	1 \%/hr	&	1.51	&	unknown	&	BU14	&	Assuming P$_{\text{rot}}$ = 2x observation length	\\
	&		&		&		&		&		&	yes	&	0.996$\mu$m	&	11 \%	&	11	&	unknown	&	GO08	&		\\
	&		&		&		&		&		&	yes	&	1.008$\mu$m	&	5 \%	&	5	&	unknown	&	GO08	&		\\
	&		&		&		&		&		&	yes	&	1.065$\mu$m	&	14 \%	&	14	&	unknown	&	GO08	&		\\
	&		&		&		&		&		&	yes	&	3.6$\mu$m	&	0.81 $\pm$ 0.08 \%	&	0.81	&	no	&	ME15	&		\\
	&		&		&		&		&		&	yes	&	4.5$\mu$m	&	1.4 $\pm$ 0.3 \%	&	1.4	&	no	&	ME15	&		\\
2MASS J08283419-1309198	&	L2	&	SC02	&	<63	&	4.9	&	AN13	&	---	&	---	&	---	&	---	&	---	&	---	&	No info	\\
2MASS J08300825+4828482	&	L9.5	&	SC14	&	<87	&	4.9	&	AN13	&	---	&	---	&	---	&	---	&	---	&	---	&	No info	\\
2MASS J08354256-0819237	&	L6.5	&	SC14	&	<14.7	&	4 -- 8	&	This work	&	yes	&	I	&	0.016	&	0.09	&	yes	&	KO04a	&	I = 17.6	\\
	&		&		&		&		&		&	yes	&	Js	&	1.3 $\pm$ 0.2 \%	&	1.3	&	yes	&	RA14b	&		\\
	&		&		&		&		&		&	no	&	Rc	&	0	&	0	&	---	&	KO05	&		\\
	&		&		&		&		&		&	yes	&	Js	&	1.6 $\pm$ 0.5 \%	&	1.6	&	yes	&	WI14	&		\\
	&		&		&		&		&		&	no	&	0.9$\mu$m	&	0	&	0	&	---	&	SC17	&		\\
	&		&		&		&		&		&	no	&	0.96$\mu$m	&	0	&	0	&	---	&	SC17	&		\\
	&		&		&		&		&		&	no	&	1.02$\mu$m	&	0	&	0	&	---	&	SC17	&		\\
	&		&		&		&		&		&	no	&	1.08$\mu$m	&	0	&	0	&	---	&	SC17	&		\\
	&		&		&		&		&		&	no	&	1.15$\mu$m	&	0	&	0	&	---	&	SC17	&		\\
	&		&		&		&		&		&	no	&	1.21$\mu$m	&	0	&	0	&	---	&	SC17	&		\\
	&		&		&		&		&		&	no	&	1.27$\mu$m	&	0	&	0	&	---	&	SC17	&		\\
	&		&		&		&		&		&	no	&	1.33$\mu$m	&	0	&	0	&	---	&	SC17	&		\\
	&		&		&		&		&		&	no	&	1.39$\mu$m	&	0	&	0	&	---	&	SC17	&		\\
	&		&		&		&		&		&	no	&	1.45$\mu$m	&	0	&	0	&	---	&	SC17	&		\\
	&		&		&		&		&		&	no	&	1.51$\mu$m	&	0	&	0	&	---	&	SC17	&		\\
	&		&		&		&		&		&	no	&	1.58$\mu$m	&	0	&	0	&	---	&	SC17	&		\\
	&		&		&		&		&		&	no	&	1.64$\mu$m	&	0	&	0	&	---	&	SC17	&		\\
	&		&		&		&		&		&	no	&	1.70$\mu$m	&	0	&	0	&	---	&	SC17	&		\\
	&		&		&		&		&		&	no	&	1.76$\mu$m	&	0	&	0	&	---	&	SC17	&		\\
	&		&		&		&		&		&	no	&	1.82$\mu$m	&	0	&	0	&	---	&	SC17	&		\\
	&		&		&		&		&		&	no	&	1.88$\mu$m	&	0	&	0	&	---	&	SC17	&		\\
	&		&		&		&		&		&	no	&	1.94$\mu$m	&	0	&	0	&	---	&	SC17	&		\\
	&		&		&		&		&		&	no	&	2.01$\mu$m	&	0	&	0	&	---	&	SC17	&		\\
	&		&		&		&		&		&	no	&	2.07$\mu$m	&	0	&	0	&	---	&	SC17	&		\\
	&		&		&		&		&		&	no	&	2.13$\mu$m	&	0	&	0	&	---	&	SC17	&		\\
	&		&		&		&		&		&	no	&	2.19$\mu$m	&	0	&	0	&	---	&	SC17	&		\\
	&		&		&		&		&		&	no	&	2.25$\mu$m	&	0	&	0	&	---	&	SC17	&		\\
	&		&		&		&		&		&	no	&	2.31$\mu$m	&	0	&	0	&	---	&	SC17	&		\\
	&		&		&		&		&		&	no	&	2.37$\mu$m	&	0	&	0	&	---	&	SC17	&		\\
2MASS J08503593+1057156 	&	L6.5+L8.5	&	DU12	&	<1302	&	4.75	&	RW13	&	---	&	---	&	---	&	---	&	---	&	---	&	No info	\\
2MASS J08575849+5708514	&	L7	&	SC14	&	<51	&	4.9	&	AN13	&	---	&	---	&	---	&	---	&	---	&	---	&	No info	\\
2MASS J09002367+2539345	&	L6.7V	&	BG14	&	<1906	&	4.75	&	RW16	&	---	&	---	&	---	&	---	&	---	&	---	&	No info	\\
2MASS J09083803+5032088	&	L8	&	SC14	&	<111	&	4.9	&	AN13	&	no	&	JH	&	0	&	0	&	---	&	BU14	&		\\
2MASS J09121469+1459396	&	L8.5+L7.5	&	DU12	&	<1473	&	4.75	&	RW13	&	---	&	---	&	---	&	---	&	---	&	---	&	No info	\\
2MASS J09130320+1841501	&	L3	&	FA09	&	<102	&	8.46	&	ML12	&	no	&	I	&	0	&	0	&	---	&	BM99	&		\\
2MASS J09230861+2340152 	&	L2.3V	&	BG14	&	<4785	&	4.75	&	RW16	&	---	&	---	&	---	&	---	&	---	&	---	&	No info	\\
2MASS J09293364+3429527	&	L8V	&	KI00	&	<42	&	8.46	&	BE06	&	---	&	---	&	---	&	---	&	---	&	---	&	No info	\\
2MASS J10101480-0406499	&	L6	&	CR03	&	<47.4	&	4 -- 8	&	This work	&	yes	&	J	&	3.6 $\pm$ 0.4 \%	&	3.6	&	yes	&	RA14b	&		\\
	&		&		&		&		&		&	yes	&	J	&	5.1 $\pm$ 1.1 \%	&	5.1	&	yes	&	WI14	&		\\
2MASS J10292165+1626526	&	L2.5	&	KI00	&	<33	&	8.46	&	ML12	&	no	&	I	&	0	&	0	&	---	&	MA07	&		\\
	&		&		&		&		&		&	no	&	R	&	0	&	0	&	---	&	MA07	&		\\
	&		&		&		&		&		&	no	&	I	&	0	&	0	&	---	&	GE02	&		\\
2MASS J10430758+2225236	&	L8.5	&	SC14	&	9.5	&	8 -- 12	&	KAO18	&	---	&	---	&	---	&	---	&	---	&	---	&	No info	\\
2MASS J10433508+1213149	&	L9	&	SC14	&	<12.6	&	4 -- 8	&	This work	&	yes	&	3.6$\mu$m	&	1.54 $\pm$ 0.15 \%	&	1.54	&	no	&	ME15	&		\\
	&		&		&		&		&		&	yes	&	4.5$\mu$m	&	1.2 $\pm$ 0.2 \%	&	1.2	&	no	&	ME15	&		\\
2MASS J10491891-5319100	&	L7.5+T0.5	&	BUR13	&	<15	&	5.5	&	OS15	&	yes	&	0.91 (0.75 − 1.10) $\mu$m	&	11 $\pm$ 1 \%	&	11	&	yes	&	GI13 	&	B	\\
	&		&		&		&		&		&	yes	&	0.91 (0.75 − 1.10)$\mu$m	&	6 $\pm$ 1 \%	&	6	&	no	&	GI13 	&	B	\\
	&		&		&		&		&		&	yes	&	0.89 (0.81 − 1.06)$\mu$m	&	7 $\pm$ 0.5 \%	&	7-11	&	no	&	BI13	&	B	\\
	&		&		&		&		&		&	no	&	1.23 (1.10 − 1.40)$\mu$m	&	0	&	0	&	---	&	BI13	&	B	\\
	&		&		&		&		&		&	yes	&	1.63 (1.50 − 1.80)$\mu$m	&	13 $\pm$ 2 \%	&	13	&	no	&	BI13	&	B	\\
	&		&		&		&		&		&	yes	&	2.16 (1.99 − 2.35)$\mu$m	&	10 $\pm$ 2 \%	&	10	&	no	&	BI13	&	B	\\
	&		&		&		&		&		&	yes	&	0.91 (0.75 − 1.10) $\mu$m	&	5 $\pm$ 1 \%	&	5	&	yes	&	BUR14 	&	B	\\
	&		&		&		&		&		&	yes	&	1.00 − 1.30 $\mu$m	&	7.5 \%	&	7.5	&	no	&	BUR14 	&	B	\\
	&		&		&		&		&		&	yes	&	1.1 - 1.6 $\mu$m	&	7 -- 11 \%	&	11	&	no	&	BU15a	&	A - no, aperiodic; B - yes	\\
	&		&		&		&		&		&	yes	&	0.8--1.15 $\mu$m 	&	9.3 \%	&	9.3	&	no	&	BU15b	&	B	\\
	&		&		&		&		&		&	yes	&	0.8--1.15 $\mu$m 	&	4.5 \%	&	4.5	&	no	&	BU15b	&	A	\\
2MASS J10584787-1548172	&	L3V	&	KI99	&	<10.5	&	4 -- 8	&	This work	&	yes	&	3.6$\mu$m	&	0.39 $\pm$ 0.04 \%	&	0.39	&	yes	&	ME15	&		\\
	&		&		&		&		&		&	no	&	4.5$\mu$m	&	0	&	0	&	---	&	ME15	&		\\
	&		&		&		&		&		&	no	&	I	&	0	&	0	&	---	&	KO13	&		\\
	&		&		&		&		&		&	no	&	R	&	0	&	0	&	---	&	KO13	&		\\
	&		&		&		&		&		&	yes	&	3.6$\mu$m	&	0.388 $\pm$ 0.043 \%	&	0.388	&	yes	&	HE13	&		\\
	&		&		&		&		&		&	yes	&	4.5$\mu$m	&	0.090 $\pm$ 0.056 \%	&	0.09	&	yes	&	HE13	&		\\
	&		&		&		&		&		&	yes	&	J	&	0.843 $\pm$ 0.098 \%	&	0.843	&	yes	&	HE13	&		\\
2MASS J11040127+1959217	&	L4	&	SC14	&	<1381	&	4.75	&	RW16		---	&	---	&	---	&	---	&	---	&	---	&	No info	\\
2MASS J11122567+3548131 	&	L4.5+L6	&	DU12	&	<1473	&	4.75	&	RW13	&	---	&	---	&	---	&	---	&	---	&	---	&	No info	\\
2MASS J11463449+2230527	&	L3+L4	&	PB08	&	<1146	&	4.75	&	RW13	&	no	&	I	&	0	&	0	&	---	&	BM99	&		\\
	&		&		&		&		&		&	yes	&	I	&	0.015	&	0.09	&	yes	&	BM01	&	I = 17.62	\\
	&		&		&		&		&		&	no	&	I	&	0	&	0	&	---	&	GE02	&		\\
	&		&		&		&		&		&	no	&	I	&	0	&	0	&	---	&	CL02	&		\\
2MASS J12035812+0015500	&	L5.0V	&	BG14	&	<63	&	8.46	&	ML12	&	---	&	---	&	---	&	---	&	---	&	---	&	No info	\\
2MASS J12195156+3128497	&	L9	&	SC14	&	<14.1	&	4 -- 8 	&	This work	&	yes	&	J	&	3 -- 6 \%/hr	&	5	&	unknown	&	BU14	&	Assuming P$_{\text{rot}}$ = 2x observation length	\\
	&		&		&		&		&		&	no	&	H	&	0	&	0	&	---	&	BU14	&		\\
2MASS J12281523-1547342	&	L5.5+L5.5	&	DU12	&	<87	&	8.46	&	BE02		no	&	Ic	&	0	&	0	&	---	&	KO13	&		\\
	&		&		&		&		&		&	no	&	Js	&	0	&	0	&	---	&	WI14	&		\\
2MASS J12560183-1257276 b	&	L7	&	GA15a	&	<9	&	8 -- 12	&	GU18		---	&	---	&	---	&	---	&	---	&	---	&	No info	\\
2MASS J13054019-2541059	&	L2+L3.5	&	KO13	&	<27.6	&	8.3	&	KR99		yes	&	857 nm	&	1.1 \%	&	1.1	&	yes	&	CL02	&		\\
	&		&		&		&		&		&	yes	&	I	&	1.2 \%	&	1.2	&	yes	&	CL03	&		\\
	&		&		&		&		&		&	yes	&	g'	&	0.04 \%	&	0.04	&	yes	&	LI06	&		\\
	&		&		&		&		&		&	no	&	5900 A	&	0	&	0	&	---	&	LI06	&		\\
	&		&		&		&		&		&	yes	&	I	&	0.0064	&	0.04	&	yes	&	KO13	&		\\
	&		&		&		&		&		&	yes	&	R	&	0.0067	&	0.03	&	yes	&	KO13	&	I = 16.85, R = 19.500	\\
2MASS J13153094-2649513 AB	&	L5+T7	&	KI11	&	370	&	5.5	&	BUR13	&	no	&	I	&	0	&	0	&	---	&	KO13	&		\\
	&		&		&		&		&		&	no	&	I	&	0	&	0	&	---	&	KO03	&		\\
	&		&		&		&		&		&	no	&	J	&	0	&	0	&	---	&	KH13	&		\\
	&		&		&		&		&		&	no	&	K	&	0	&	0	&	---	&	KH13	&		\\
2MASS J13285503+2114486 	&	L4.1V	&	BG14	&	<1158	&	4.75	&	RW13	&	---	&	---	&	---	&	---	&	---	&	---	&	No info	\\
2MASS J14243909+0917104	&	L4	&	LE01	&	<97	&	8.46	&	BE02	&	---	&	---	&	---	&	---	&	---	&	---	&	No info	\\
2MASS J14252798-3650229	&	L4	&	GA15b	&	<12.9	&	4 -- 8	&	This work	&	yes	&	J	&	0.6 $\pm$ 0.1 \%	&	0.6	&	yes	&	RA14a	&		\\
	&		&		&		&		&		&	yes	&	Js	&	0.7 $\pm$ 0.3 \%	&	0.7	&	yes	&	VO18	&		\\
2MASS J14460061+0024519	&	L4.2V	&	BG14	&	<1098	&	4.75	&	RW13	&	---	&	---	&	---	&	---	&	---	&	---	&	No info	\\
2MASS J15065441+1321060	&	L3	&	SC14	&	<78	&	8.46	&	ML12	&	no	&	I	&	0	&	0	&	---	&	MA07	&		\\
	&		&		&		&		&		&	no	&	R	&	0	&	0	&	---	&	MA07	&		\\
	&		&		&		&		&		&	no	&	I	&	0	&	0	&	---	&	GE02	&		\\
2MASS J15074769-1627386	&	L5V	&	KI00	&	<36.6	&	9	&	LY16		no	&	Ic	&	0	&	0	&	---	&	KO03	&		\\
	&		&		&		&		&		&	no	&	Ic	&	0	&	0	&	---	&	KO13	&		\\
	&		&		&		&		&		&	no	&	Js	&	0	&	0	&	---	&	WI14	&		\\
	&		&		&		&		&		&	yes	&	1.4$\mu$m	&	40 \%	&	40	&	unknown	&	YA15	&		\\
	&		&		&		&		&		&	no	&	Js	&	0	&	0	&	---	&	RA14b	&		\\
	&		&		&		&		&		&	yes	&	3.6$\mu$m	&	0.53 $\pm$ 0.11 \%	&	0.53	&	no	&	ME15	&		\\
	&		&		&		&		&		&	yes	&	4.5$\mu$m	&	0.45 $\pm$ 0.09 \%	&	0.45	&	no	&	ME15	&		\\
2MASS J15150083+4847416	&	L6.5	&	CR03	&	<27	&	8.46	&	BE06	&	no	&	1.1--1.7 $\mu$m 	&	0	&	0	&	---	&	BU14	&		\\
2MASS J15232263+3014562 	&	L8V	&	KI00	&	<45	&	8.46	&	BE06	&	---	&	---	&	---	&	---	&	---	&	---	&	No info	\\
2MASS J16154255+4953211	&	L4gamma	&	CR18	&	<9.0	&	4 -- 8	&	This work	&	yes	&	3.6$\mu$m	&	0.9 $\pm$ 0.2 \%	&	0.9	&	unknown	&	ME15	&		\\
	&		&		&		&		&		&	no	&	4.5$\mu$m	&	0	&	0	&	---	&	ME15	&		\\
	&		&		&		&		&		&	no	&	J	&	0	&	0	&	---	&	VO18	&		\\
2MASS J16154416+3559005	&	L3V	&	KI00	&	<75	&	8.46	&	ML12	&	no	&	I	&	0	&	0	&	---	&	GE02	&		\\
2MASS J16322911+1904407	&	L8	&	SC14	&	<10.8	&	4 -- 8	&	This work	&	no	&	J	&	0	&	0	&	---	&	BU14	&		\\
	&		&		&		&		&		&	no	&	H	&	0	&	0	&	---	&	BU14	&		\\
	&		&		&		&		&		&	no	&	Js	&	0	&	0	&	---	&	WI14	&		\\
	&		&		&		&		&		&	yes	&	3.6$\mu$m	&	0.42 $\pm$ 0.08 \%	&	0.42	&	yes	&	ME15	&		\\
	&		&		&		&		&		&	yes	&	4.5$\mu$m	&	0.5 $\pm$ 0.3 \%	&	0.5	&	yes	&	ME15	&		\\
2MASS J17072343-0558249	&	M9+L3	&	RE08	&	<48	&	8.46	&	BE06	&	---	&	---	&	---	&	---	&	---	&	---	&	No info	\\
2MASS J17114573+2232044	&	L5.0+T5.5	&	BUR10	&	<11.4	&	4 -- 8	&	This work	&	yes	&	J	&	0.206 $\pm$ 0.041	&	1.21	&	yes	&	KH13	&		\\
	&		&		&		&		&		&	yes	&	K'	&	1.186 $\pm$ 0.083	&	8.05	&	yes	&	KH13	&		\\
	&		&		&		&		&		&	no	&	J	&	0	&	0	&	---	&	BU14	&		\\
2MASS J17210390+3344160	&	L5.3:V	&	BG14	&	<48	&	8.46	&	BE06	&	yes	&	3.6$\mu$m	&	0.33 $\pm$ 0.07 \%	&	0.33	&	yes	&	ME15	&		\\
	&		&		&		&		&		&	no	&	4.5$\mu$m	&	0	&	0	&	---	&	ME15	&		\\
2MASS J17281150+3948593	&	L5+L6.5	&	GE14	&	<54	&	8.46	&	BE06	&	---	&	---	&	---	&	---	&	---	&	---	&	No info	\\
2MASS J17502484-0016151	&	L5	&	KO17	&	185	&	4 -- 8	&	This work	&	no	&	I	&	0	&	0	&	---	&	KO13	&		\\
	&		&		&		&		&		&	no	&	R	&	0	&	0	&	---	&	KO13	&		\\
	&		&		&		&		&		&	yes	&	J	&	0.75 \%/hr	&	0.5	&	unknown	&	BU14	&	Assuming P$_{\text{rot}}$ = 2x observation length	\\
	&		&		&		&		&		&	no	&	H	&	0	&	0	&	---	&	BU14	&		\\
	&		&		&		&		&		&	no	&	J	&	0	&	0	&	---	&	RA14a	&		\\
2MASS J18212815+1414010	&	L5	&	SC14	&	<12.9	&	4 -- 8	&	This work	&	no	&	I	&	0	&	0	&	---	&	KO13	&		\\
	&		&		&		&		&		&	no	&	R	&	0	&	0	&	---	&	KO13	&		\\
	&		&		&		&		&		&	yes	&	3.6$\mu$m	&	0.54 $\pm$ 0.05 \%	&	0.54	&	no	&	ME15	&		\\
	&		&		&		&		&		&	yes	&	4.5$\mu$m	&	0.71 $\pm$ 0.14 \%	&	0.71	&	no	&	ME15	&		\\
	&		&		&		&		&		&	yes	&	1.1-1.7$\mu$m	&	1.77 $\pm$ 0.11 \%	&	1.77	&	unknown	&	YA15	&		\\
	&		&		&		&		&		&	yes	&	1.4$\mu$m	&	1.54 $\pm$ 0.21 \%	&	1.54	&	unknown	&	YA15	&		\\
	&		&		&		&		&		&	yes	&	0.90	&	3 \%	&	3	&	yes	&	SC17	&		\\
	&		&		&		&		&		&	yes	&	0.96	&	2.5 \%	&	2.5	&	yes	&	SC17	&		\\
	&		&		&		&		&		&	yes	&	1.02	&	2.5 \%	&	2.5	&	yes	&	SC17	&		\\
	&		&		&		&		&		&	yes	&	1.08	&	1.8 \%	&	1.8	&	yes	&	SC17	&		\\
	&		&		&		&		&		&	yes	&	1.15	&	1.8 \%	&	1.8	&	yes	&	SC17	&		\\
	&		&		&		&		&		&	yes	&	1.21	&	1.5 \%	&	1.5	&	yes	&	SC17	&		\\
	&		&		&		&		&		&	yes	&	1.27	&	1.3 \%	&	1.3	&	yes	&	SC17	&		\\
	&		&		&		&		&		&	yes	&	1.33	&	1.3 \%	&	1.3	&	yes	&	SC17	&		\\
	&		&		&		&		&		&	yes	&	1.39	&	1.3 \%	&	1.3	&	yes	&	SC17	&		\\
	&		&		&		&		&		&	yes	&	1.45	&	1.3 \%	&	1.3	&	yes	&	SC17	&		\\
	&		&		&		&		&		&	yes	&	1.51	&	0.9 \%	&	0.9	&	yes	&	SC17	&		\\
	&		&		&		&		&		&	yes	&	1.58	&	0.8 \%	&	0.8	&	yes	&	SC17	&		\\
	&		&		&		&		&		&	yes	&	1.64	&	0.7 \%	&	0.7	&	yes	&	SC17	&		\\
	&		&		&		&		&		&	yes	&	1.70	&	0.5 \%	&	0.5	&	yes	&	SC17	&		\\
	&		&		&		&		&		&	no	&	1.76	&	0	&	0	&	---	&	SC17	&		\\
	&		&		&		&		&		&	no	&	1.82	&	0	&	0	&	---	&	SC17	&		\\
	&		&		&		&		&		&	no	&	1.88	&	0	&	0	&	---	&	SC17	&		\\
	&		&		&		&		&		&	yes	&	1.94	&	0.6 \%	&	0.6	&	yes	&	SC17	&		\\
	&		&		&		&		&		&	no	&	2.01	&	0	&	0	&	---	&	SC17	&		\\
	&		&		&		&		&		&	no	&	2.07	&	0	&	0	&	---	&	SC17	&		\\
	&		&		&		&		&		&	no	&	2.13	&	0	&	0	&	---	&	SC17	&		\\
	&		&		&		&		&		&	no	&	2.19	&	0	&	0	&	---	&	SC17	&		\\
	&		&		&		&		&		&	no	&	2.25	&	0	&	0	&	---	&	SC17	&		\\
	&		&		&		&		&		&	no	&	2.31	&	0	&	0	&	---	&	SC17	&		\\
	&		&		&		&		&		&	no	&	2.37	&	0	&	0	&	---	&	SC17	&		\\
2MASS J18410861+3117279	&	L4Vpec	&	KI00	&	<3696	&	4.75	&	RW13	&	---	&	---	&	---	&	---	&	---	&	---	&	No info	\\
2MASS J21011544+1756586	&	L7+L8	&	DU12	&	<3172	&	4.75	&	RW13	&	---	&	---	&	---	&	---	&	---	&	---	&	No info	\\
2MASS J21041491-1037369	&	L2	&	SC14	&	<24	&	8.46	&	BE06	&	no	&	I	&	0	&	0	&	---	&	KO13	&		\\
	&		&		&		&		&		&	no	&	JHK	&	0	&	0	&	---	&	KT05	&		\\
2MASS J21481633+4003594	&	L7	&	SC14	&	<9.6	&	4 -- 8	&	This work	&	no	&	J	&	0	&	0	&	---	&	KH13	&		\\
	&		&		&		&		&		&	yes	&	3.6$\mu$m	&	1.33 $\pm$ 0.07  \%	&	1.33	&	yes	&	ME15	&		\\
	&		&		&		&		&		&	yes	&	4.5$\mu$m	&	1.03 $\pm$ 0.1  \%	&	1.03	&	yes	&	ME15	&		\\
2MASS J22244381-0158521	&	L4.5V	&	KI00	&	<33	&	8.46	&	BE06	&	yes	&	I	&	0.083	&	0.46	&	no	&	GE02	&	I  = 18.0	\\
	&		&		&		&		&		&	no	&	JHK	&	0	&	0	&	---	&	KO04b	&		\\
	&		&		&		&		&		&	no	&	JHK	&	0	&	0	&	---	&	KT05	&		\\
	&		&		&		&		&		&	no	&	3.6$\mu$m	&	0	&	0	&	---	&	ME15	&		\\
	&		&		&		&		&		&	no	&	4.5$\mu$m	&	0	&	0	&	---	&	ME15	&		\\
2MASS J22521073-1730134	&	L4.5+T3.5	&	DU12	&	<30	&	8.46	&	BE06	&	no	&	Js	&	0	&	0	&	---	&	WI14	&		\\
	&		&		&		&		&		&	no	&	I	&	0	&	0	&	---	&	KO13	&		\\
\enddata
\tablerefs{
(AN13)		\citet{antonova2013};
(BE02)		\citet{Berger2002};
(BE05)		\citet{Berger2005};
(BE06)		\citet{berger2006};
(BG14)		\citet{Bardalez2014};
(BI13)		\citet{Biller2013};
(BL08)		\citet{Blake2008};
(BM01)		\citet{bailer-jones2001};
(BM99)		\citet{Bailer-Jones1999};
(BU14)		\citet{buenzli2014};
(BU15a)		\citet{Buenzli2015a};
(BU15b)		\citet{Buenzli2015b};
(BUR10)		\citet{Burgasser2010}; 
(BUR13)		\citet{Burgasser2013};
(BUR14)		\citet{Burgasser2014};
(CA13)		\citet{Castro2013};
(CL02)		\citet{Clarke2002};
(CL03)		\citet{Clarke2003};
(CL08)		\citet{clarke2008};
(CR03)		\citet{cruz2003};
(CR16)		\citet{Croll2016};
(CR18)		\citet{Cruz2018};
(DU12)		\citet{Dupuy2012};
(DU17)		\citet{Dupuy2017}; 
(EBB03)		\citet{enoch2003};
(FA09)		\citet{Faherty2009};
(GA15a)		\citet{Gagne2015a};
(GA15b)		\citet{Gagne2015b};
(GE02)		\citet{Gelino2002};
(GE14)		\citet{Gelino2014};
(GI13)		\citet{Gillon2013};
(GI16)		\citet{gizis2016};
(GO08)		\citet{Goldman2008};
(GU09)		\citet{Guenther2009};
(GU18)		\citet{Guirado2018};
(HA13)		\citet{Harding2013};
(HE13)		\citet{Heinze2013};
(KAO16)		\citet{Kao2016};
(KAO18)		\citet{kao2018};
(KH13)		\citet{Khandrika2013};
(KI00)		\citet{kirkpatrick2000};
(KI11)		\citet{Kirkpatrick2011};
(KI99)		\citet{kirkpatrick1999};
(KO03)		\citet{Koen2003};
(KO04a)		\citet{koen2004a};
(KO04b)		\citet{Koen2004b};
(KO05)		\citet{Koen2005};
(KO13)		\citet{Koen2013};
(KO17)		\citet{Koen2017};
(KR99)		\citet{Krishnamurthi1999};
(KT05)		\citet{Koenetal2005};
(LA07)		\citet{Lane2007};
(LE01)		\citet{Leggett2001};
(LI06)		\citet{Littlefair2006};
(LY16)		\citet{Lynch2016};
(MA07)		\citet{Maiti2007};
(ML12)		\citet{McLean2012};
(ME15)		\citet{Metchev2015};
(OS15)		\citet{Osten2015};
(PB08)		\citet{Phan-Bao2008};
(RA14a)		\citet{Radigan2014a};
(RA14b)		\citet{Radigan2014};
(RE00)		\citet{Reid2000}; 
(RE06)		\citet{Reid2006};
(RE08)		\citet{reid2008};
(RW13)		\citet{Route2013};
(RW16)		\citet{Route2016};
(SC02)		\citet{Scholz2002};
(SC07)		\citet{Schmidt2007};
(SC14)		\citet{schneider2014};
(SC15)		\citet{schmidt2015};
(SC17)		\citet{Schlawin2017};
(VO18)		\citet{Vos2018};
(WI14)		\citet{wilson2014};
(YA15)		\citet{yang2015};
}
\tablenotetext{a}{The fluxes reported are at frequencies between 4 -- 12 GHz.}
\end{deluxetable}
\end{longrotatetable}

\end{document}